\documentclass[11pt]{article}
\usepackage{graphics}
\usepackage{lscape,epsfig}
\usepackage{amsmath}
\usepackage{amsfonts}
\usepackage{amssymb}
\usepackage{natbib}
\usepackage{enumerate}
\usepackage[title,titletoc,toc]{appendix}
\usepackage{color}
\usepackage{tikz}
\usepackage[normalem]{ulem}
\usepackage{mathtools}
\usepackage{authblk}
\usepackage{hyperref}
\usepackage{rotating}

\usetikzlibrary{arrows,positioning,shapes.geometric}
\definecolor{red}   {RGB}{180,0,0}
\definecolor{gray10}{rgb}{0.1,0.1,0.1}
\definecolor{gray20}{rgb}{0.2,0.2,0.2}
\definecolor{gray30}{rgb}{0.3,0.3,0.3}
\definecolor{gray40}{rgb}{0.4,0.4,0.4}
\definecolor{gray50}{rgb}{0.5,0.5,0.5}
\definecolor{gray60}{rgb}{0.6,0.6,0.6}
\definecolor{gray80}{rgb}{0.8,0.8,0.8}
\definecolor{gray90}{rgb}{0.9,0.9,.9}
\definecolor{gray95}{rgb}{0.95,0.95,.95}
\definecolor{gray96}{rgb}{0.96,0.96,.96}
\definecolor{sgGreen} {RGB}{20, 180, 50}
\newcommand{\dashedrightarrow}[1][2pt]{  \settowidth{\@tempdima}{$\rightarrow$}\rightarrow  \makebox[-\@tempdima]{\hskip-1.5ex\color{white}\rule[0.5ex]{#1}{1pt}}  \phantom{\rightarrow}}
\newtheorem{innercustomgeneric}{\customgenericname}
\providecommand{\customgenericname}{}
\newcommand{\newcustomtheorem}[2]{%
  \newenvironment{#1}[1]
  {%
   \renewcommand\customgenericname{#2}%
   \renewcommand\theinnercustomgeneric{##1}%
   \innercustomgeneric
  }
  {\endinnercustomgeneric}
}

\newcustomtheorem{customthm}{Theorem}
\newcustomtheorem{customlemma}{Lemma}
\newcustomtheorem{customcorollary}{Corollary}

\DeclarePairedDelimiterX\Basics[1](){ #1}

\def\appendix{\par \setcounter{section}{0} \def\thesection{A}}
\begin{document}

\title{A causal framework for classical statistical estimands in failure time settings with competing events}

\author[1]{Jessica G. Young\thanks{Corresponding Author: jyoung@hsph.harvard.edu}}

\author[2,3]{Mats J. Stensrud}

\author[4]{Eric J. Tchetgen Tchetgen}

\author[2,5,6]{Miguel A. Hern\'{a}n}

\affil[1]{Department of Population Medicine, Harvard Medical School and Harvard Pilgrim Health Care Institute, MA, USA}
\affil[2]{Department of Epidemiology Harvard T.H. Chan School of Public Health, MA, USA}
\affil[3]{Department of Biostatistics, Oslo Centre for Biostatistics and Epidemiology, University of Oslo, Norway}
\affil[4]{Department of Statistics, The Wharton School, University of Pennsylvania, PA, USA}
\affil[5]{Department of Biostatistics Harvard T.H. Chan School of Public Health, MA, USA}
\affil[6]{Harvard-MIT Division of Health Sciences and Technology, MA, USA}

\date{}
\setcounter{Maxaffil}{0}
\renewcommand\Affilfont{\itshape\small}

\maketitle

\begin{abstract}
In failure-time settings, a competing event is any event that makes it impossible for the event of interest to occur. For example, cardiovascular disease death is a competing event for prostate cancer death because an individual cannot die of prostate cancer once he has died of cardiovascular disease.  Various statistical estimands have been defined as possible targets of inference in the classical competing risks literature.  Many reviews have described these
statistical estimands and their estimating procedures with recommendations about their use. However, this previous work has not used a formal framework for
characterizing causal effects and their identifying conditions, which makes it difficult to interpret effect estimates and assess recommendations regarding analytic choices. Here we use a counterfactual framework to explicitly define each of these classical estimands. 
We clarify that, depending on whether competing events are defined as censoring events, contrasts of
risks can define a \textsl{total effect} of the treatment on the event of
interest, or a \textsl{direct effect} of the treatment on the event of
interest not mediated through the competing event.
In contrast, regardless of whether competing events are defined as censoring events, counterfactual hazard contrasts cannot
generally be interpreted as causal effects. We illustrate how identifying assumptions for all of these
counterfactual estimands can be represented in causal diagrams %
 in which competing events are depicted as
time-varying covariates.  We present an application of these ideas to data from a randomized trial designed to estimate the effect of estrogen therapy on prostate cancer mortality.\end{abstract}

\section{Introduction}\label{intro}

In failure-time settings, a competing event is any event
that makes it impossible for the event of interest to occur. For example,
death from cardiovascular disease is a competing event for death from
prostate cancer because an individual cannot die of prostate cancer once he
has died of cardiovascular disease. Because competing events cannot be
prevented by design, they can occur in both randomized and nonrandomized
studies.

Various statistical estimands have been defined as possible targets of
inference in the classical literature on competing risks in failure-time
settings. As recently summarized by \cite{geskus}, these include the
so-called \textsl{marginal cumulative incidence} (alternatively, \textsl{net
risk}), \textsl{cause-specific cumulative incidence}
(alternatively, \textsl{%
subdistribution function} or \textsl{crude risk}), \textsl{marginal hazard}, \textsl{%
subdistribution hazard}, and \textsl{cause-specific hazard}%
\citep{geskus,finegray,kalbprentice}. Many reviews describe these
statistical estimands and their estimating procedures \cite%
{gooley,pintilie,pintilie1,pintilie2,andersencr,laucr,edwards,geskus}.  Early authors also considered the interpretation of these estimands\citep{chiang1,chiang2} with others
providing recommendations about their use \citep{laucr,austincirc,geskus,latouche}. However, this previous work has not used a formal framework for
characterizing causal effects and their identifying conditions, which makes
it difficult to interpret the effect estimates from these procedures and to assess recommendations
regarding analytic choices.

Here we use a counterfactual framework \citep{robinsfail,causalitypearl} to
explicitly define each of these classical statistical estimands, which
results in two different counterfactual definitions of risk and three
different counterfactual definitions of hazard. We clarify that contrasts of
risks can define a \textsl{total effect} of the treatment on the event of
interest, or a \textsl{direct effect} of the treatment on the event of the
interest that is not mediated through the competing event. A key distinction
between these definitions of causal effect is whether competing events are
defined as censoring events. In contrast, regardless of whether competing events are defined as censoring events, contrasts of hazards cannot
generally be interpreted as causal effects. We also show how, contrary to
previous claims \citep{leskolau}, identifying assumptions for all of these
counterfactual estimands can be represented in causal diagrams %
\citep{pearldag,swigs} in which the competing event is depicted as a
time-varying covariate.

The manuscript is structured as follows. In Section \ref{obsdata}, we
describe the longitudinal observed data structure of interest. In Section \ref%
{defnocomprisk}, we give counterfactual definitions of classical statistical
estimands in failure-time settings and define causal effects when competing
events do not exist. In Section \ref{estimand}, we give counterfactual
definitions of classical estimands and definitions of total and direct
effects when competing events exist. We also consider the relation between the choice of estimand and the definition of a censoring event. In Section \ref{identification}, we
provide assumptions under which these counterfactual estimands may be
identified, illustrate how causal diagrams may be used to evaluate these
assumptions, and show that the identifying functions constitute special
cases of Robins's g-formula\citep{robinsfail}. In
Section \ref{choosing} we consider how to choose between different definitions
of causal effect when competing events exist.  In Section \ref{stat}, we
outline various estimators of total and direct effects based on various
algebraically equivalent representations of the g-formula.   In Section \ref%
{analysis}, we present an application of these ideas to data from a
randomized trial designed to estimate the effect of estrogen therapy on
prostate cancer mortality. In Section \ref%
{Discussion}, we provide a discussion, including brief consideration of so-called \textsl{cross-world counterfactual} alternatives to
our definition of a direct effect when competing events exist.

\section{Observed data structure}\label{obsdata}
Consider a randomized trial in which each of $i=1,\ldots ,n$
individuals with prostate cancer are randomly assigned to either treatment $%
A=1$ (assignment to estrogen therapy) or $A=0$ (assignment to placebo) at baseline. Individuals are
assumed independent and identically distributed and thus we suppress an
individual-specific $i$ subscript. Let $k=0,\ldots ,K+1$ denote equally
spaced follow-up intervals (e.g., months) with interval $k=0$ corresponding
to baseline and interval $k=K+1$ corresponding to a maximum follow-up of
interest (e.g., 60 months post-baseline) selected less than or equal to the maximum possible follow-up time (i.e., the administrative end of the study, beyond which no data on any individual are available). Let $Y_{k}$ and $D_k$
denote indicators of the event of interest (e.g., death from prostate
cancer) and a competing event (e.g., death from
cardiovascular disease) \textsl{by} interval $k$, respectively. By definition, $D_{0}\equiv Y_{0}\equiv 0$ because the study population is restricted to those who have not yet experienced the event of interest or the competing event at
baseline.

For $k>0$, let $L_{k}$ denote a vector of time-varying individual
characteristics updated by $k$ (e.g., indicator of a nonfatal
cardiovascular event) with baseline
covariates $L_{0}$ (e.g. baseline physical activity level, baseline age,
history of cardiovascular disease) measured before assignment to treatment $%
A $. We assume the temporal ordering $(D_{k},Y_{k},L_{k})$ within each
follow-up interval $k>0$. For simplicity, we also assume that all variables
are measured without error. Note that when time-varying covariates are not measured for all individuals in all intervals $k>0$, $L_k$ may contain last measured values of patient characteristics along with the time since an updated measurement \citep{obsplans}.

We denote the history of a random variable using overbars, e.g., $\overline{Y%
}_{k}=(Y_{0},\ldots ,Y_{k})$ is the history of the event of interest through
interval $k$. We denote the future of a random variable through the
follow-up of interest using underbars, e.g., $\underline{Y}%
_{k+1}=(Y_{k+1},\ldots ,Y_{K+1})$. Notably, if an
individual is known to experience the competing event by interval $k>0$
without history of the event of interest ($Y_{k-1}=0,D_{k}=1$) then all
future indicators for the event of interest $(\underline{Y}_{k})$ will be
observed (known) and deterministically zero because, by definition, individuals who
experience a competing event can never subsequently experience the event of
interest. We assume no loss to follow-up until Section \ref{censdef}.

\section{Counterfactual estimands when competing events do not exist}\label{defnocomprisk}

Suppose we are interested in the causal effect of a point treatment $%
A$ on the event of interest. In this section we begin with the simplified
case in which competing events do not exist, i.e. $\overline{D}_{K+1}\equiv 
\overline{0}$, which will occur when death from any cause is the event of
interest. In the next section, we consider the more general case in which
competing events exist, which will occur when death from prostate cancer is
the event of interest and death from another cause (e.g., cardiovascular
disease) is a competing event.

To define the causal effect when competing events do not exist, we first
need to define the counterfactual (or potential) outcome variables $%
Y_{k+1}^{a}$ for $a=0,1$ and $k=0,...K$. For each individual, $Y_{k+1}^{a}$
is the indicator of the event of interest by interval $k+1$ if the
individual, possibly contrary to fact, had been assigned to $A=a$. We can then define the counterfactual
risk of the event of interest by $k+1$ had all individuals in the population
been assigned $A=a$ as 
\begin{equation}
\Pr [Y_{k+1}^{a}=1],  \label{eqn:risk}
\end{equation}
$k=0,\ldots ,K$. The average
causal effect of treatment $A$ on the event of interest by $k+1$ can, in turn, be
defined by 
\begin{equation}
\Pr [Y_{k+1}^{a=1}=1]-\Pr[Y_{k+1}^{a=0}=1]\label{eqn:effect}.
\end{equation}
such that the treatment $A$ has a nonnull average causal effect on the event of
interest by $k+1$ if and only if $\Pr [Y_{k+1}^{a=1}=1]\neq \Pr
[Y_{k+1}^{a=0}=1]$. More generally, the population-level causal effect of treatment $A$ on the event of interest by $k+1$ can be defined as $\Pr [Y_{k+1}^{a=1}=1]\mbox{ vs. }[\Pr [Y_{k+1}^{a=0}=1]$, even on a non-additive (e.g. ratio) scale.  

We can analogously define the discrete-time hazard of the event of interest
in interval $k+1$ under $a$ as 
\begin{equation}
\Pr [Y_{k+1}^{a}=1|Y_{k}^{a}=0]  \label{eqn:hazard}
\end{equation}%
We will refer to \eqref{eqn:hazard} as a \textsl{discrete-time hazard%
} regardless of whether the underlying counterfactual failure time is
discrete or continuous. That is, defining $T^{a}$ as the
counterfactual time to failure from the event of interest under an
intervention that sets $A$ to $a$, we can
equivalently write \eqref{eqn:hazard} as $\Pr [T^{a}\in
(t_{k},t_{k+1}]|T^{a}>t_{k}]$ with interval $k$ defined by $%
(t_{k},t_{k+1}]$. This is a discrete-time hazard when $T^{a}$
is discrete with support at $t_{k+1}$, $k=0,\ldots ,K$. The limit of this
function as interval-length approaches zero divided by interval length is
the continuous-time hazard function when $T^{a}$ is
continuous.

Unlike the risk \eqref{eqn:risk}, the hazard \eqref{eqn:hazard} is
conditional on survival to $k$, which may be affected by treatment $A$ (for $%
k>0$). Therefore, $\Pr [Y_{k+1}^{a=1}=1|Y_{k}^{a=1}=0]\neq \Pr [Y_{k+1}^{a=0}=1|Y_{k}^{a=0}=0]$ does not necessarily imply
that $A$ has a nonnull causal effect at $k+1$. The hazards at $k+1$ may
differ just because of differences in individuals who survive until $k$
under $a=1$ versus $a=0$ due to treatment effects before $k$ %
\citep{Hernanfail,hazards}.  For this reason, unlike the contrast in
counterfactual risks \eqref{eqn:effect}, we generally cannot interpret
contrasts in counterfactual hazards 
\begin{equation}
\Pr [Y_{k+1}^{a=1}=1|Y_{k}^{a=1}=0]\mbox{ vs. }\Pr [Y_{k+1}^{a=0}=1|Y_{k}^{a=0}=0]  \label{eqn:hazardcontrast}
\end{equation}%
as \textsl{causal effects} even though they may be precisely defined
contrasts of counterfactual quantities and may even be identifiable from the
study data \citep{flanders}. See also Section \ref{identifyhazards}.

\section{Counterfactual estimands when competing events exist}\label{estimand}
When competing events exist, the counterfactual
outcomes---and therefore risks and hazards---can be defined in different
ways. In this section, we describe various counterfactual definitions of
risk and hazard of the event of interest when competing events exist and map
these to estimands that have been defined in the classical statistical
literature (see Table 1).  We also give an interpretation of contrasts in
these counterfactual estimands under different interventions on treatment $A$.

\begin{sidewaystable}[!htp]
\caption{Definitions, descriptions and terminology for different counterfactual risks and hazards of the event of interest when competing events exist under an intervention that sets $A$ to $a$.$^{*}$ }
\label{sens}
\begin{center}
\begin{tabular}{lcc}
\hline
\noalign{\smallskip} Definition & Description$^{**}$ & Terminology from the statistical literature$^{***}$
 \\ \hline
$\Pr [Y_{k+1}^{a,\overline{d}=\overline{0}}=1] $ &  risk under elimination of competing events& marginal cumulative incidence, \\
&&net risk  \\ 
$\Pr [Y_{k+1}^{a}=1]$  & risk without elimination of competing events & subdistribution function, \\ 
&&cause-specific cumulative incidence  \\ 
&&crude risk\\
$\Pr [Y_{k+1}^{a,\overline{d}=\overline{0}}=1|Y_{k}^{a,\overline{d}=\overline{0}}=0] $ &  hazard under elimination of competing events & marginal hazard  \\ 
$\Pr [Y_{k+1}^{a}=1|Y_{k}^{a}=0] $& hazard without elimination of competing events  & subdistribution hazard   \\ 
$\Pr [Y_{k+1}^{a}=1|D_{k+1}^{a}=Y_{k}^{a}=0]$ &  hazard conditioned on competing events & cause-specific hazard  \\ 
\hline
\end{tabular}%
\end{center}
\par
{\footnotesize{$^{*}$ When loss to follow-up may occur for some individuals in the study population by some $k+1$, $k=0,\ldots,K$, we will index all estimands by $\overline{c}=0$, an additional intervention that eliminates loss to follow-up (see Section \ref{censdef}). 
}}
\par
{\footnotesize{$^{**}$ Similar descriptions of risk were given in the early statistical literature on competing events but without definition of counterfactuals \citep{chiang1}. 
}}
\par
{\footnotesize{$^{***}$ Based on Table 1.1 of a recent textbook by \cite{geskus}. 
}}
\par
{\footnotesize \ }
\end{sidewaystable}

\subsection{Direct effects}
For $k=0,\ldots,K$, consider the counterfactual outcome $Y_{k+1}^{a,\overline{d}=\overline{0}}$ where $\overline{d}=\overline{0}$
represents a hypothetical intervention that eliminates the competing event.
Then the counterfactual
risk by $k+1$ under $a$ 
\begin{equation}
\Pr [Y_{k+1}^{a,\overline{d}=\overline{0}}=1]  \label{eqn:risk1}
\end{equation}%
is the risk that would have been observed if all individuals had been
assigned to treatment $a$ and we had somehow eliminated competing events. This risk under elimination of competing
events has been referred to as the \textsl{marginal cumulative incidence} or 
\textsl{net risk} in the statistics literature \citep{geskus}.

Under this definition of the counterfactual outcome, the average causal
effect of treatment $A$ on the event of interest by $k+1$ is 
\begin{equation}
\Pr [Y_{k+1}^{a=1,\overline{d}=\overline{0}}=1]-\Pr
[Y_{k+1}^{a=0,\overline{d}=\overline{0}}=1]  \label{eqn:RD1}
\end{equation}%
The contrast \eqref{eqn:RD1} is a (controlled) direct effect %
\citep{rgmediation} which quantifies the treatment's effect on
the event of interest not mediated by competing events. The causal directed acyclic graph (DAG) \citep{pearldag} in Figure \ref{dagimperfectfor14} depicts an underlying data generating assumption for two arbitrary follow-up times under which the competing event may mediate the treatment's effect on the event of interest; e.g. via the causal path $A\rightarrow D_k\rightarrow Y_k $ or the causal path $A\rightarrow D_k\rightarrow D_{k+1}\rightarrow Y_{k+1} $.  Note that the arrow from $D_k$ to $Y_{k}$ (or $D_{k+1}$ to $Y_{k+1}$) must be included because, by definition, if $D_{k}=1$ then $Y_k=0$.

\subsection{Total effects}
Suppose we consider the alternative counterfactual outcome $Y_{k+1}^{a}$, $k=0,\ldots ,K$, which does not entail any
hypothetical intervention on competing events. Note that there are two types of individuals
with $Y_{k+1}^{a}=0$: those who survive both the event of
interest and the competing event (e.g., do not die of any cause) through $%
k+1 $ and those who experience a competing event (e.g., die from
cardiovascular disease) by $k+1$. The counterfactual risk by $k+1$ under $a$ 
\begin{equation}
\Pr [Y_{k+1}^{a}=1]  \label{eqn:risk2}
\end{equation}%
is the risk that would have been observed if all individuals had been
assigned to treatment $a$ without elimination of competing events. This risk without elimination of competing events %
\eqref{eqn:risk2} can be alternatively represented as $\Pr [T^{a}\leq t_{k+1},J^{a}=1]$ where $T^{a}$ is the
counterfactual time to either the event of interest ($J^{a}=1$%
) or a competing event ($J^{a}=2$), whichever comes first.
Equivalence between $\Pr [T^{a}\leq t_{k+1},J^{a}=1]$ and the risk without elimination of competing events follows by $%
Y_{k+1}^{a}=I(T^{a}\leq t_{k+1},J^{a}=1)$, with $I(\cdot )$ the indicator function. The risk without
elimination of competing events has been referred to as the \textsl{%
subdistribution function},  \textsl{%
cause-specific cumulative incidence function}, or \textsl{crude risk} for
cause $J=1$ at $t_{k+1}$\citep{kalbprentice,finegray,geskus}. Note that both the risk
under elimination of competing events \eqref{eqn:risk1} \textsl{and} the
risk without elimination of competing events \eqref{eqn:risk2} are
\textquotedblleft marginal\textquotedblright\ (population-level) risks but they correspond to the marginal distributions of different counterfactual outcomes.

Under this definition of the counterfactual outcome, the average causal
effect of treatment $A$ on the event of interest by $k+1$ is 
\begin{equation}
\Pr [Y_{k+1}^{a=1}=1]-\Pr [Y_{k+1}^{a=0}=1].  \label{eqn:RD2}
\end{equation}%
The contrast \eqref{eqn:RD2} quantifies the total effect of treatment
on the event of interest through all causal pathways between treatment and
the event of interest, including those possibly mediated by the competing
event (see Figure \ref{dagimperfectfor14}).

As we discuss in Section \ref{choosing}, it will be useful to define the risk
of the competing event by $k+1$ under $a$ 
\begin{equation}
\Pr [D_{k+1}^{a}=1]  \label{eqn:risk3}
\end{equation}%
where $D_{k+1}^{a}$ is an indicator of the competing event by $k+1$ had, possibly
contrary to fact, an individual received $A=a$. When, as in our example, the original event of interest is also a competing
event for the original competing event (that is, when the event of interest
and the competing event are \textsl{mutually competing events}), the risk %
\eqref{eqn:risk3} has been referred to as the \textsl{subdistribution
function}, \textsl{cause-specific
cumulative incidence function}, or \textsl{crude risk} for cause $J=2$ at $%
t_{k+1}$ \citep{kalbprentice,finegray,geskus}. Our presentation is generally agnostic as to whether the data structure corresponds to this mutually competing events setting or a so-called \textsl{semi-competing risk} setting \citep{semicr}, noting when any distinction is needed.  A semi-competing risk setting would occur, for example, were the event of interest prostate cancer \textsl{diagnosis}, which is not a competing event for death from other causes.

A definition of the average causal effect of treatment $A$ on the competing
event by $k+1$ is then 
\begin{equation}
\Pr [D_{k+1}^{a=1}=1]-\Pr [D_{k+1}^{a=0}=1]  \label{eqn:RD4}
\end{equation}%
which quantifies the total effect of treatment on the competing event
through all causal pathways between treatment and the competing event,
including those possibly mediated by the event of interest.

Another common estimand in competing risks settings is the result of
redefining the event of interest as a composite outcome. In our example,
this would be equivalent to changing the event of interest from prostate
cancer death to death from any cause, which is precisely the case in which
competing events do not exist and the estimands of Section \ref%
{defnocomprisk} apply. The risk of this composite outcome by $k+1$ under $a$
is $\Pr [Y_{k+1}^{a}=1\mbox{ or }D_{k+1}^{a}=1]$ and the average causal effect of treatment $A$ on this
composite outcome by $k+1$ is 
\begin{equation}
\Pr [Y_{k+1}^{a=1}=1\mbox{ or }D_{k+1}^{a=1}=1]-\Pr [Y_{k+1}^{a=0}=1%
\mbox{ or }D_{k+1}^{a=0}=1]  \label{eqn:RD3}
\end{equation}%
which quantifies the total effect of treatment on the composite outcome by $%
k+1$ through all causal pathways. It is straightforward to see that, when the event of interest and the competing event are mutually competing events, the effect \eqref{eqn:RD3} is simply the sum of \eqref{eqn:RD2} and \eqref{eqn:RD4}.

\subsection{Counterfactual hazards}
Using the various counterfactual outcome definitions above, various
discrete-time hazards can be defined. First, the \textsl{hazard under
elimination of competing events}
\begin{equation}
\Pr [Y_{k+1}^{a,\overline{d}=\overline{0}}=1|Y_{k}^{a,\overline{d}=\overline{0}}=0]  \label{eqn:hazard1}
\end{equation}%
is the hazard of the event of interest at $k+1$ if all individuals had been
assigned to treatment $a$ and we had somehow eliminated competing events. For $T^{a,\overline{d}=0}$, the corresponding
counterfactual failure time, we can equivalently write \eqref{eqn:hazard1} as $\Pr
[T^{a,\overline{d}=0}\in (t_{k},t_{k+1}]|T^{a,\overline{d}=0}>t_{k}]$. The hazard under elimination of competing events
has been referred to as the \textsl{marginal hazard} \citep{geskus}.

Second, the \textsl{hazard without elimination of competing events}%
\begin{equation}
\Pr [Y_{k+1}^{a}=1|Y_{k}^{a}=0]  \label{eqn:hazard2}
\end{equation}%
is the hazard of the event of interest at $k+1$ had all individuals been
assigned to treatment $a$ without elimination of competing events. The \textquotedblleft risk set\textquotedblright\
of individuals at $k$, i.e, those with $Y_{k}^{a}=0 $, in this case is comprised by (i) those who have experienced neither
the event of interest nor the competing event, and (ii) those who have not
experienced the event of interest but have experienced the competing event
by $k$. The hazard without elimination of competing events %
\eqref{eqn:hazard2} can alternatively be represented as $\Pr [T^{a}\in (t_{k},t_{k+1}],J^{a}=1|(T^{a}>t_{k}%
\mbox{ or }\{T^{a}\leq t_{k}\mbox{ and }J^{a}\neq 1\})]$, which follows by $Y_{k+1}^{a}=I(T^{a}\leq t_{k+1},J^{a}=1)$ and $I(Y_{k}^{a}=0)=I(\{Y_{k}^{a}=0,D_{k}^{a}=0\}\mbox{ or }\{Y_{k}^{a}=0,D_{k}^{a}=1\})=I(T^{a}>t_{k}\mbox{ or }\{T^{a}\leq t_{k}\mbox{ and }J^{a}\neq 1\})$.  This quantity has been referred to as the \textsl{subdistribution hazard for
cause $J=1$} \citep{finegray}.

A third definition is the hazard of the event
of interest at $k+1$ among those who have not previously
experienced the competing event. When $A$ is set to $a$, this \textsl{hazard conditioned on competing events}
can be written as 
\begin{equation}
\Pr [Y_{k+1}^{a}=1|D_{k+1}^{a}=Y_{k}^{a}=0]  \label{eqn:hazard3}
\end{equation}%
or $\Pr [T^{a}\in (t_{k},t_{k+1}],J^{a}=1|T^{a}>t_{k}]$. This quantity has been called the \textsl{cause-specific hazard} for cause $J=1$ \citep{geskus,andersencr}.

By these three different definitions of the hazard, we might consider three
different contrasts in counterfactual hazards (e.g. hazard ratios) under $a=1$ versus $a=0$: a contrast in the hazards
under elimination of competing events 
\begin{equation}
\Pr [Y_{k+1}^{a=1,\overline{d}=\overline{0}}=1|Y_{k}^{a=1,\overline{d}=\overline{0}}=0]\mbox{ vs. }\Pr [Y_{k+1}^{a=0,\overline{d}=\overline{0}}=1|Y_{k}^{a=0,\overline{d}=\overline{0}}=0],  \label{eqn:hazard1contrast}
\end{equation}%
a contrast in the hazards without elimination of competing events
(subdistribution hazards) 
\begin{equation}
\Pr [Y_{k+1}^{a=1}=1|Y_{k}^{a=1}=0]\mbox{ vs. }\Pr [Y_{k+1}^{a=0}=1|Y_{k}^{a=0}=0],  \label{eqn:hazard2contrast}
\end{equation}%
or a contrast in the hazards conditioned on competing events
(cause-specific hazards) 
\begin{equation}
\Pr [Y_{k+1}^{a=1}=1|D_{k+1}^{a=1}=Y_{k}^{a=1}=0]\mbox{ vs. }\Pr
[Y_{k+1}^{a=0}=1|D_{k+1}^{a=0}=Y_{k}^{a=0}=0]
\label{eqn:hazard3contrast}
\end{equation}%
As in the case where competing events do not exist, none of these three contrasts in counterfactual hazards
can in general be interpreted as a causal effect \citep{Hernanfail,hazards}.
Unlike risks, hazards may differ just because of differences in individuals
who survive until $k$ under $a=1$ versus $a=0$ due to treatment effects
before $k$ (also see Section \ref{identifyhazards}).

Finally, in later sections we will reference the hazard of the competing event
itself at $k+1$ among those who have not previously experienced the event of
interest under $a$. This coincides with the \textsl{cause-specific hazard }for
cause $J=2$ or 
\begin{equation}
\Pr [D_{k+1}^{a}=1|D_{k}^{a}=Y_{k}^{a}=0].  \label{eqn:hazard4}
\end{equation}%
which is alternatively written $\Pr [T^{a}\in
(t_{k},t_{k+1}],J^{a}=2|T^{a}>t_{k}]$.

\section{Definition of a censoring event}\label{censdef}
We now relax the assumption that there is no loss to follow-up in the study of Section \ref{obsdata}.  Define $C_{k+1}$ as an indicator of loss to follow-up (i.e., end of study observation) by interval $k+1$ for $k=0,\ldots,K$, and  $C_0\equiv 0$. Assume the temporal order $(C_{k+1},D_{k+1},Y_{k+1})$ so that, if an individual is lost to follow-up by $%
k+1$ ($C_{k+1}=1$), all future indicators of both the event of interest
(all components of $\underline{Y}_{k+1}$) and the competing event (all
components of $\underline{D}_{k+1})$ will not be observed, that is, will be missing. For simplicity,we assume that, if $C_{k+1}=0$ then $(\overline{D}_k,\overline{Y}_k)$ is fully observed.

Loss to follow-up is always understood as a form of \textsl{censoring}.  However, it is often debated whether competing events are censoring events. In this section, we give a definition of a \textsl{censoring event} which clarifies that the choice to define a competing event as a censoring event depends on the choice of estimand. 

To do so, let the \textsl{counterfactual outcomes of interest under $a$ } be those counterfactual outcomes by $k+1$, $k=0,\ldots,K$, on which the investigator-chosen estimand depends.  When the estimand depends on counterfactual outcomes indexed \textsl{only} by an intervention on $a$, then we will say $Y^{a}_{k+1}$ is the counterfactual outcome of interest under $a$.  Otherwise, define $V_{k+1}$ as a vector whose elements are indicators of the occurrence of a selected set of one or more events in the study by $k+1$, $k=0,\ldots,K$ such that no one in the study population experiences those events before baseline.  When the estimand depends on counterfactual outcomes indexed not only by an intervention on $a$, but also by an intervention $\overline{v}=0$ that (somehow) eliminates those events throughout the follow-up, then $Y_{k+1}^{a,\overline{v}=0}$ is the counterfactual outcome of interest under $a$.  

We now give the following definition of a \textsl{censoring event}: 
\begin{align}
    & \text{A censoring event is any event occurring in the study by } k+1,k=0,\ldots,K, \nonumber \\
    &\text{that ensures the values of all future counterfactual outcomes of interest under } a \nonumber \\ 
    & \text{are unknown even for an individual receiving the intervention } a. \label{def: definition1} 
\end{align}

By this definition, when the counterfactual outcomes of interest under $a$ are $Y_{k+1}^{a,\overline{v}=0}$ for some choice of $V_{k+1}$, all elements of $V_{k+1}$ are censoring events.  For example, suppose the investigator chooses to estimate either the direct effect \eqref{eqn:RD1} or the hazard contrast \eqref{eqn:hazard1contrast}.  These estimands contrast functionals of the counterfactual outcomes  $Y_{k+1}^{a,\overline{d}=0}$ under different levels of $a$.  For these estimands, $D_{k+1}$ is a censoring 
event because, for individuals who experience a competing event by $k+1$%
, their future status on the event of interest under $a$ and an intervention that eliminates competing events $\underline{Y}_{k+1}^{a,\overline{d}=0}$ is unknown. Alternatively, suppose the investigator chooses to estimate the total effect \eqref{eqn:RD2}, the (subdistribution) hazard contrast \eqref{eqn:hazard2contrast} or  the (cause-specific) hazard contrast \eqref{eqn:hazard3contrast}. These estimands contrast functionals of the counterfactual outcomes  $Y_{k+1}^{a}$ under different levels of $a$.  For these estimands, competing events are
\textsl{not} censoring events because, by definition of a competing event, individuals with $%
(Y_{k}^{a}=0,D_{k+1}^{a}=1)$ have known future outcomes $\underline{Y}_{k+1}^{a}=\underline{0}$ as they can never experience the event of interest.  

Note that loss to follow-up $C_{k+1}$ meets our definition of a censoring event even if the counterfactual outcomes of interest under $a$ are not indexed by an intervention that ``eliminates loss to follow-up'', $\overline{c}=0$. Such counterfactual outcomes can be understood as the outcomes under $a$ and the natural values of $\overline{C}_{k+1}$ under $a$ \citep{swigs}, e.g. $Y_{k+1}^{a}\equiv Y_{k+1}^{a,\overline{C}^{a}_{k+1}}$  (also see Appendix \ref{appasub}).  However, when loss to follow-up occurs in the study, the identification results of the next section are only generally applicable to estimands indexed by this additional intervention $\overline{c}=0$ \citep{causalbook}; therefore, we restrict consideration to counterfactual outcomes under $a$ of the form $Y_{k+1}^{a,\overline{v}=0}$ with $C_{k+1}$ always an element of $V_{k+1}$.  In some settings, it may be reasonable to assume that loss to follow-up does not affect future events. Under this additional constraint, our identification results will apply to estimands with or without the $\overline{c}=0$ index because, in this case, a counterfactual risk or hazard indexed by an intervention $\overline{c}=0$ will equal its counterpart without this intervention (e.g., we will have $Y_{k+1}^{a,\overline{c}=0}=Y_{k+1}^{a}$ for all individuals in the study population for all $k$).  

Unlike definitions of \textsl{right-censoring} of a failure time in the classical survival analysis literature, our definition only refers to an event that precludes knowledge/observation of future status on time-varying indicators of failure from the event of interest only through $K+1$.  It does not refer to a \textsl{failure time}. Still, our definition is consistent with classical definitions of right-censoring in the sense that, if the indicators $\underline{Y}^{a,\overline{v}=0}_{k+1}$ are unobserved due to censoring events $V_{k+1}$, then the counterfactual failure time $T^{a,\overline{v}=0}$ will also be unobserved. On the other hand, reaching the administrative end of the study without failure, resulting in so-called 
\textsl{administrative} right-censoring of the failure time $T^{a,\overline{v}=0}$%
, does \textsl{not} result in failure to observe the indicator values $\underline{Y}_{k+1}^{a,\overline{%
v}=0}$ at any $k\leq K+1$ because, as stated in Section \ref{obsdata}, we have restricted consideration to $K+1$ at or before the administrative study end. Thus, surviving all causes of failure through the administrative end of the study does \textsl{not} constitute a censoring event by definition \eqref{def: definition1}. For completeness, we note that causal inference with data containing \textsl{left-censored} failure times (occurring before study enrollment) will require particularly strong assumptions.  Unlike right-censored data, left-censored data can be avoided by, as in Section \ref{obsdata}, restricting the study population to those alive and at risk of all events at the start of the study. 

Finally, suppose the estimand is chosen such that only loss to follow-up, and not competing events, is a censoring event by the definition \eqref{def: definition1}.   In this case, an individual who experiences a competing event by $k+1$ without prior censoring cannot subsequently experience a censoring event.  For example, once an individual dies of cardiovascular disease, we know he cannot subsequently die of prostate cancer at any future time and there is no subsequent event he can experience that prevents this knowledge.   Importantly, nowhere in this presentation do we require the assumption that there exists a \textsl{potential censoring time} \citep{finegray} for an individual observed to fail during the study period.

\section{Identification of estimands when competing events exist }\label{identification}

We now provide conditions under which the various counterfactual estimands of Section \ref%
{estimand}, all now additionally indexed by interventions $\overline{c}=0$, can be identified when the data described in Section \ref{obsdata}
are available and loss to follow-up may also occur. The nature of these conditions depends on whether competing
events are censoring events by the definition \eqref{def: definition1}. Without loss of generality, we
consider identification of risks only by $K+1$.

\subsection{Direct effects 
\label{casei}}

\begin{figure}[tbp]
\centering
\includegraphics[scale=1]{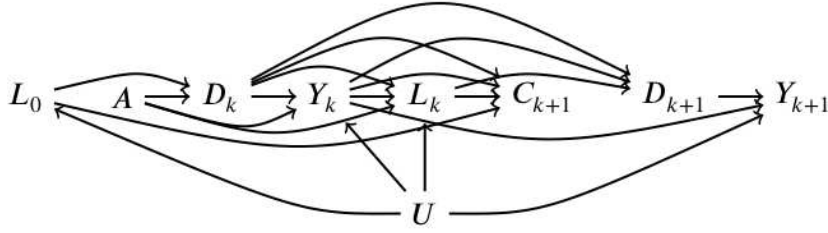}
\caption{A causal DAG representing observed data generating assumptions
under which (i) competing events may mediate the effect of treatment $A$ on the event of interest and (ii) exchangeability \eqref{eqn:C1} holds such that the direct
effect \eqref{eqn:RD1} may be identified. }
\label{dagimperfectfor14}
\end{figure}

To identify the risk under elimination of competing events, and in turn the
direct effect \eqref{eqn:RD1}, we must make untestable assumptions. For each 
$k=0,\ldots ,K$, consider the following three identifying assumptions:

\begin{enumerate}

\item Exchangeability 1:
\begin{equation*}
\overline{Y}_{K+1}^{a,\overline{c}=\overline{d}=0}\perp\!\!\!\perp A|L_0,
\end{equation*}
\begin{equation}
\underline{Y}_{k+1}^{a,\overline{c}=\overline{d}=0}\perp\!\!\!\perp (C_{k+1},D_{k+1})|%
\overline{L}_k=\overline{l}_k,\overline{Y}_k=\overline{C}_k=\overline{D}%
_k=0,A=a  \label{eqn:C1}
\end{equation}
where $\overline{l}_{k}$ is some realization of $\overline{L}_{k}$. This
assumption requires that, in addition to the baseline observed treatment, at
each follow-up time, conditional on the measured past, all forms of
censoring are independent of future counterfactual outcomes had everyone
followed $A=a$ and censoring were eliminated. Because loss to follow-up and
competing events cannot be randomly assigned by an investigator in practice,
this condition will not hold by design, even in an experiment in which $A$
is randomized.

The causal DAG in
Figure \ref{dagimperfectfor14} depicts a data generation process under which exchangeability \eqref{eqn:C1} holds 
by the absence of unblocked
backdoor paths \citep{pearldag,causalbook} between (i) $A$ and both $Y_k$ and $Y_{k+1}$ conditional on $L_0$ and (ii) $D_{k}$ and both $Y_k$ and $Y_{k+1}$ conditional on $A$ and $L_0$; $C_{k+1}$
and $Y_{k+1}$ conditional on $L_{k}$, $Y_k$, $D_{k}$, $A$ and $L_0$; and $D_{k+1}$
and $Y_{k+1}$ conditional on $C_{k+1}$, $L_{k}$, $Y_k$, $D_{k}$, $A$ and $L_0$.  In
Figure \ref{dagimperfectfor14}, (ii) is guaranteed by the absence of arrows
from $U$, an unmeasured risk factor for the event of interest, into $D_{k}$, $C_{k+1}$ and $D_{k+1}$. Note that we have omitted
other arrows on the graph (e.g. an arrow from $L_{k}$ to $Y_{k+1}$) to reduce
clutter as adding any missing arrows from past into future measured
variables will still preserve (i) and (ii). For interested readers, in Appendix \ref{appasub} we illustrate an alternative approach to graphical representation of exchangeability using Single World Intervention Graphs (SWIGS)\citep{swigs} which allows explicit representation of counterfactuals on the graph.  This approach, in turn, makes more explicit the role of the target counterfactual estimand in graphical evaluation of exchangeability.

\item Positivity 1:
\begin{align}
&f_{A,\overline{L}_{k},C_k,D_{k},Y_k}(a,\overline{l}_{k},0,0,0)\neq0\implies\nonumber\\
&\Pr[C_{k+1}=0,D_{k+1}=0|\overline{L}_k=\overline{l}_{k},C_k=D_k=Y_k=0,A=a]>0\mbox{ w.p.1}. \label{eqn:positivity1}
\end{align}
where $f_{A,\overline{L}_{k},C_k,D_{k},Y_k}(a,\overline{l}_{k},0,0,0)$ is the joint density of  $(A,\overline{L}_k,C_k,D_k,Y_k)$ evaluated at $(a,\overline{l}_{k},0,0,0)$. This assumption requires that, for any possibly observed level of treatment and covariate history amongst those remaining uncensored (here free of competing events and loss to follow-up) and free of the  event of interest through $k$,  some individuals continue to remain uncensored through $k+1$.

\item Consistency 1:
\begin{align}
&\mbox{If } A=a \mbox{ and }\overline{C}_{k+1}=\overline{D}_{k+1}=0\mbox{,} \nonumber\\
&\mbox{ then } \overline{L}_{k+1}=\overline{L}_{k+1}^{a,\overline{c}=\overline{d}=0} \mbox{ and } \overline{Y}_{k+1}=\overline{Y}_{k+1}^{a,\overline{c}=\overline{d}=0} \label{eqn:consistency1}
\end{align}
This assumption requires that, if an individual has data consistent with the
interventions indexing the counterfactual outcome through $k+1$, then her
observed outcomes and covariates through $k+1$ equal her counterfactual
outcomes and covariates under that intervention. The consistency assumption
requires well-defined interventions, which is particularly problematic when,
as here, the estimand implies an unspecified intervention to eliminate death
from other causes \citep{welldefined2,welldefined1,waterkill}.
\end{enumerate}

Under conditions \eqref{eqn:C1}, \eqref{eqn:positivity1} and %
\eqref{eqn:consistency1}, the risk under elimination of competing events by $%
K+1$, $\Pr [Y_{K+1}^{a,\overline{c}=\overline{d}=\overline{0}}=1]$, is
identified by the following function of the observed data: 
\begin{align}
\sum_{\overline{l}_{K}}\sum_{k=0}^{K}& \Pr [Y_{k+1}=1|\overline{L}_{k}=%
\overline{l}_{k},\overline{C}_{k+1}=\overline{D}_{k+1}=\overline{Y}%
_{k}=0,A=a]\times  \notag \\
& \prod_{j=0}^{k}\{\Pr [Y_{j}=0|\overline{L}_{j-1}=\overline{l}_{j-1},%
\overline{C}_{j}=\overline{D}_{j}=\overline{Y}_{j-1}=0,A=a]\times  \notag \\
& f(l_{j}|\overline{l}_{j-1},\overline{C}_{j}=\overline{D}_{j}=\overline{Y}%
_{j}=0,a)\}  \label{eqn:gform1}
\end{align}%
where $\Pr [Y_{k+1}=1|\overline{L}_{k}=\overline{l}_{k},\overline{C}_{k+1}=%
\overline{D}_{k+1}=\overline{Y}_{k}=0,A=a]$ is the observed discrete-time hazard at $%
k+1$ of the event of interest conditional on treatment and covariate history
among those still free of the competing event and not lost to follow-up, and $%
f(l_{j}|\overline{l}_{j-1},\overline{C}_{j}=\overline{D}_{j}=\overline{Y}%
_{j}=0,a)$ is the conditional density of $L_{j}$. For $j=0$, $f(l_{j}|%
\overline{l}_{j-1},\overline{C}_{j}=\overline{D}_{j}=\overline{Y}%
_{j}=0,a)\equiv f(l_{0})$.

We say that expression \eqref{eqn:gform1} is the \textsl{g-formula} for the risk under elimination of competing events $\Pr [Y_{K+1}^{a,%
\overline{c}=\overline{d}=\overline{0}}=1]$ \citep{robinsfail}.  The proof of equivalence between the g-formula \eqref{eqn:gform1} and $\Pr [Y_{K+1}^{a,\overline{c}=\overline{d}=\overline{0}}=1]$ under conditions \eqref{eqn:C1}, \eqref{eqn:positivity1} and %
\eqref{eqn:consistency1} was given by Robins [\citeyear{robinsfail,jamieidentifiability}] and is reviewed in Appendix \ref{appb}. The g-formula \eqref{eqn:gform1} has several algebraically equivalent representations.
For example, we can
equivalently write \eqref{eqn:gform1} using the following inverse probability
weighted (IPW) representation 
\begin{equation}
\sum_{k=0}^{K}h_{k}(a)\prod_{j=0}^{k-1}\left[1-h_{j}(a)\right]  \label{eqn:ipw1}
\end{equation}
where 
\begin{equation}
h_{k}(a)=\frac{\mbox{E}\lbrack Y_{k+1}(1-Y_{k})W_{k}|A=a]}{\mbox{E}\lbrack
(1-Y_{k})W_{k}]A=a]}  \label{eqn:h1}
\end{equation}
and $W_{k}=W^{C}_k\times W^{D}_k$ with 
\begin{equation*}
W^{C}_k=\prod_{j=0}^{k}\frac{I(C_{j+1}=0)}{\Pr[C_{j+1}=0|%
\overline{L}_j,\overline{C}_{j}=\overline{D}_{j}=\overline{Y}%
_{j}=0,A=a]}  
\end{equation*}%
and 
\begin{equation*}
W^{D}_k=\prod_{j=0}^{k}\frac{I(D_{j+1}=0)}{\Pr[D_{j+1}=0|%
\overline{L}_j,\overline{C}_{j+1}=\overline{D}_{j}=\overline{Y%
}_{j}=0,A=a]}  
\end{equation*}%
where $\mbox{E}[\cdot ]$ denotes expectation and the
denominators of the weights $W_{k}^{C}$ and $W_{k}^{D}$ denote the
probabilities of remaining free of each type of censoring (loss to follow-up
and competing events, respectively) by $k+1$ conditional on measured
history. See Appendix \ref{appb}. Our ability to represent the g-formula in
different yet algebraically equivalent ways has implications for choices in
estimating this function in practice, as discussed in Section \ref{stat}.

We note that, in many randomized trials, including the trial analyzed in
Section \ref{analysis}, only baseline covariates $L_0$ are measured. In this
case, in order to identify causal treatment effects, a stronger version of
exchangeability \eqref{eqn:C1} must hold with $\overline{L}_k$ replaced only
by $L_0$ for all $k$. This stronger version of exchangeability requires that censoring at any time during the follow-up is independent of changing values of prognostic factors during the follow-up, often an implausible assumption, particularly when competing events are defined as censoring events.

\subsection{Total effects\label{caseii}}
To identify the risk without elimination of competing events, and in turn the
total effect \eqref{eqn:RD2}, we must also make untestable assumptions.
However, because only losses to follow-up (and not competing events) are
censoring events for total effects, the assumptions required to identify
total effects are weaker than those required to identify direct effects.

\begin{figure}[tbp]
\centering
\includegraphics[scale=1]{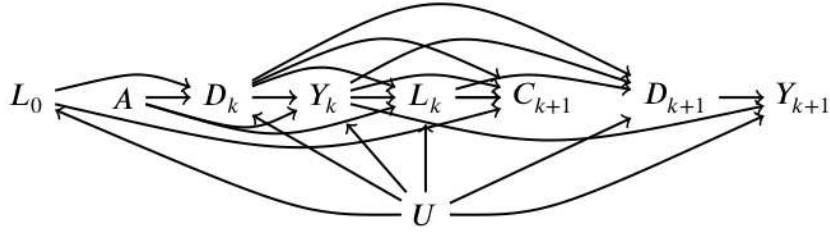}
\caption{A causal DAG representing observed data generating assumptions under which exchangeability \eqref{eqn:C2} holds and the total effect of treatment $A$ on the event of interest \eqref{eqn:RD2} may be identified.  }
\label{dag3}
\end{figure}

Specifically,  for each $k=0,\ldots ,K$, consider the following alternative
versions of exchangeability, positivity and consistency:
\begin{enumerate}
\item Exchangeability 2:
\begin{equation*}
\overline{Y}_{K+1}^{a,\overline{c}=0}\perp\!\!\!\perp A|L_0,
\end{equation*}
\begin{equation}
\underline{Y}_{k+1}^{a,\overline{c}=0}\perp\!\!\!\perp C_{k+1}|\overline{L}_k=%
\overline{l}_k,\overline{D}_k=\overline{d}_k,\overline{Y}_k=\overline{C}%
_k=0,A=a  \label{eqn:C2}
\end{equation}
where $D_{k}$ may be viewed as a covariate like those in $L_{k}$. This
assumption requires that, in addition to the baseline observed treatment, at
each follow-up time, given the measured past, censoring (here, only loss to
follow-up) is independent of future counterfactual outcomes had everyone
followed $A=a$ and censoring were eliminated. 

The causal DAG in Figure \ref%
{dag3} depicts a data generating process under which exchangeability \eqref{eqn:C2}
holds. The only difference between Figure \ref{dag3} and Figure \ref%
{dagimperfectfor14} is the former allows arrows from the
unmeasured risk factor for the event of interest $U$ into the competing
event at each time ($D_{k}$ and $D_{k+1}$).The presence of these arrows
would violate assumption \eqref{eqn:C1} rendering direct effects
unidentified. Figure \ref{dag3} is consistent with assumption  \eqref{eqn:C2} by  
 (i) the absence of any unblocked backdoor paths between $A$ and both $Y_k$ and $Y_{k+1}$ conditional on $L_0$ and (ii) the absence of such paths between
$C_{k+1}$ and $Y_{k+1}$,
conditional on $L_k$, $Y_k$, $D_{k}$, $A$ and $L_0$. The latter is
guaranteed by the lack of an arrow from $U$ into $C_{k+1}$.

\item Positivity 2:
\begin{align}
&f_{A,\overline{L}_{k},\overline{D}_k,C_k,Y_k}(a,\overline{l}_{k},\overline{d}_k,0,0)\neq0\implies\nonumber\\
&\Pr[C_{k+1}=0|\overline{L}_k=\overline{l}_{k},\overline{D}_k=\overline{d}_k,C_k=Y_k=0,A=a]>0\mbox{ w.p.1}\label{eqn:positivity2}
\end{align}
where $f_{A,\overline{L}_{k},\overline{D}_k,C_k,Y_k}(a,\overline{l}_{k},\overline{d}_k,0,0)$ is the joint density of $(A,\overline{L}_{k},\overline{D}_k,C_k,Y_k)$ evaluated at $(a,\overline{l}_{k},\overline{d}_k,0,0)$.  Note that for any $k$ such that $D_k=1$, this assumption holds by definition because, in this case, the probability of remaining uncensored by $k+1$ is 1 (individuals who fail from the competing event by $k$ without prior censoring, by definition \eqref{def: definition1}, cannot subsequently experience a censoring event).

\item Consistency 2:
\begin{align}
&\mbox{If } A=a \mbox{ and }\overline{C}_{k+1}=0\mbox{,} \nonumber\\
&\mbox{ then } \overline{L}_{k+1}=\overline{L}_{k+1}^{a,\overline{c}=0}\mbox{, }\overline{D}_{k+1}=\overline{D}_{k+1}^{a,\overline{c}=0} \mbox{ and } \overline{Y}_{k+1}=\overline{Y}_{k+1}^{a,\overline{c}=0} \label{eqn:consistency2}
\end{align}
\end{enumerate}

Under conditions \eqref{eqn:C2}, \eqref{eqn:positivity2} and %
\eqref{eqn:consistency2}, the risk without elimination of competing events
by $K+1$ $\Pr [Y_{K+1}^{a,\overline{c}=\overline{0}}=1]$ is identified by
the following g-formula: 
\begin{align}
& \sum_{\overline{l}_{K}}\sum_{\overline{d}_{K+1}}\sum_{k=0}^{K}\Pr
[Y_{k+1}=1|\overline{L}_{k}=\overline{l}_{k},\overline{D}_{k+1}=\overline{d}%
_{k+1},\overline{C}_{k+1}=\overline{Y}_{k}=0,A=a]\times  \notag \\
\prod_{j=0}^{k}\{& \Pr [Y_{j}=0|\overline{L}_{j-1}=\overline{l}_{j-1},%
\overline{D}_{j}=\overline{d}_{j},\overline{C}_{j}=\overline{Y}%
_{j-1}=0,A=a]\times  \notag \\
& \Pr[D_{j+1}=d_{j+1}|\overline{l}_{j},\overline{d}_{j},\overline{C}_{j+1}=%
\overline{Y}_{j}=0,a]\times f(l_{j}|\overline{l}_{j-1},\overline{d}_{j},%
\overline{C}_{j}=\overline{Y}_{j}=0,a)\} \label{eqn:gform2}
\end{align}
which follows directly from earlier results by Robins[\citeyear
{robinsfail,jamieidentifiability}] when, as discussed in Appendix \ref{appb}, 
$\overline{D}_{k}$ is defined as a component of $\overline{L}_{k}$.

However, because $\overline{D}_{k}$ has a deterministic relationship with
the event of interest, expression \eqref{eqn:gform2} is algebraically
equivalent to the following somewhat simplified expression \citep{taubman}
\begin{align}
& \sum_{\overline{l}_{K}}\sum_{k=0}^{K}\Pr [Y_{k+1}=1|\overline{L}_{k}=%
\overline{l}_{k},\overline{C}_{k+1}=\overline{D}_{k+1}=\overline{Y}%
_{k}=0,A=a]\times  \notag \\
\prod_{j=0}^{k}\{& \Pr [Y_{j}=0|\overline{L}_{j-1}=\overline{l}_{j-1},%
\overline{C}_{j}=\overline{D}_{j}=\overline{Y}_{j-1}=0,A=a]\times  \notag \\
& \Pr[D_{j+1}=0|\overline{l}_{j},\overline{C}_{j+1}=\overline{D}_{j}=%
\overline{Y}_{j}=0,a]\times f(l_{j}|\overline{l}_{j-1},\overline{C}_{j}=%
\overline{D}_{j}=\overline{Y}_{j}=0,a)\} \label{eqn:gform2taub}
\end{align}
where $\Pr [D_{k+1}=1|\overline{l}_{k},\overline{D}_{k}=\overline{C}_{k+1}=%
\overline{Y}_{k}=0,a]$ is the observed discrete-time hazard of the competing event in
interval $k+1$ conditional on treatment and covariate history among those
still not lost to follow-up.  This simplification results from the fact that all
terms in the sum \eqref{eqn:gform2} over $\overline{d}_{K+1}$ are zero when $%
d_{k+1}=1$ for any $k=0,\ldots ,K$.  

Analogously, there are 
different algebraically equivalent representations of \eqref{eqn:gform2taub}. One IPW representation is: 
\begin{equation}
\sum_{k=0}^{K}h^{sub}_{k}(a)\prod_{j=0}^{k-1}\left[1-h^{sub}_{j}(a)\right]
\label{eqn:ipw2first}
\end{equation}
where 
\begin{equation}
h^{sub}_{k}(a)=\frac{\mbox{E}\lbrack Y_{k+1}(1-Y_{k})W^{sub}_{k}|A=a]}{%
\mbox{E}\lbrack (1-Y_{k})W^{sub}_{k}|A=a]},  \label{eqn:H2}
\end{equation}
and 
\begin{equation*}
W^{sub}_{k}=\prod_{j=0}^{k}\frac{I(C_{j+1}=0)}{\Pr[C_{j+1}=0|\overline{D}_j,\overline{L}_j,\overline{C}_{j}=\overline{Y}%
_{j}=0,A=a]}.  
\end{equation*}%

A second algebraically equivalent IPW representation of \eqref{eqn:gform2taub} is: 
\begin{equation}
\sum_{k=0}^{K}h^{1}
_{k}(a)\{1-h^{2}_{k}(a)\}\prod_{j=0}^{k-1}[\{1-h^{1}_{j}(a)\}\{1-h^{2}_{j}(a)\}]
\label{eqn:ipw2}
\end{equation}%
where 
\begin{equation}
h^{1} _{k}(a)=\frac{\mbox{E}\lbrack\ Y_{k+1}(1-D_{k+1})(1-Y_{k})W^{C}_{k}|A=a]}{\mbox{E}\lbrack\ (1-{D_{k+1}})(1-Y_{k})W^{C}_{k}|A=a]},  \label{eqn:cs1}
\end{equation}%
\begin{equation}
h _{k}^{2}(a)=\frac{\mbox{E}\lbrack\ D_{k+1}(1-Y_{k})(1-D_{k})W^{C}_{k}|A=a]}{\mbox{E}\lbrack\ (1-{Y_{k}})(1-D_{k})W^{C}_{k}|A=a]},  \label{eqn:cs2}
\end{equation}%
and $W^{C}_{k}$ is defined as in Section \ref{casei}.  See Appendix \ref{appb}.

Note that in a trial where $A$ is randomized and no one is lost to
follow-up (there are no censoring events), then exchangeability %
\eqref{eqn:C2} is expected for $\overline{L}_k=\varnothing$. It is
straightforward to see in this case that, even when competing events exist,
the identifying function \eqref{eqn:gform2taub} for the risk without
elimination of competing events, and any of its algebraically equivalent
alternative representations, reduces simply to $\Pr[Y_{K+1}=1|A=a]$; in
turn, in this special case, the total effect on the event of interest %
\eqref{eqn:RD2} is guaranteed identified by $\Pr[Y_{K+1}=1|A=1]-%
\Pr[Y_{K+1}=1|A=0]$.

Finally, note that arguments similar to those given in this section follow
for identification of the risk of the competing event itself by $K+1$ under $%
a$ \eqref{eqn:risk3} and, in turn, the total treatment effect on the
competing event \eqref{eqn:RD4}, in data with censoring events. These
assumptions are given in Corollary 8, Appendix \ref{appb} along with the
corresponding g-formula for \eqref{eqn:risk3}.

\subsection{Counterfactual hazards and the ``hazards of hazard ratios'' revisited }\label{identifyhazards}
In
Appendix \ref{appb}, we prove the following claims:
\begin{enumerate}
    \item The same assumptions that allow identification of a direct effect on the
event of interest \eqref{eqn:RD1} also give identification of a contrast in hazards under elimination of competing events \eqref{eqn:hazard1contrast} by a contrast in the observed data function \eqref{eqn:h1}.  As discussed in Sections %
\ref{casei} and \ref{caseii}, these assumptions are consistent with 
Figure \ref{dagimperfectfor14} but fail under Figure %
\ref{dag3}.
\item The same assumptions that allow
identification of the total effect on the event of interest \eqref{eqn:RD2}
also give identification of the counterfactual contrast in subdistribution hazards %
\eqref{eqn:hazard2contrast} by a contrast in the observed data function \eqref{eqn:H2}. As in Section \ref{caseii}, these
assumptions are consistent with both Figures \ref{dagimperfectfor14} and \ref{dag3}.
\item The same assumptions that allow
identification of the total effect on the event of interest \eqref{eqn:RD2}, coupled with an additional set of assumptions that allow identification of the total effect on the competing event \eqref{eqn:RD4} (given in Corollary 8, Appendix \ref{appb}), allow identification of the counterfactual contrast in cause-specific hazards \eqref{eqn:hazard3contrast} by a contrast in the observed data function \eqref{eqn:cs1}. Similarly, these assumptions give identification of the counterfactual cause-specific hazard for the competing event itself \eqref{eqn:hazard4} by the function \eqref{eqn:cs2}. These two sets of assumptions are also consistent with both Figures \ref{dagimperfectfor14} and \ref{dag3} using graphical rules for evaluating exchangeability (the absence of unblocked backdoor paths) \citep{pearldag,swigs,causalbook}.
\end{enumerate}

It is interesting to note that the observed data function \eqref{eqn:h1} which, by claim 1 above, identifies the counterfactual hazard under elimination of competing events given the assumptions of Section \ref{casei}, is a weighted version of the 
\textsl{observed cause-specific hazards} of the event of interest
conditioned on remaining free of loss to follow-up and $A=a$.  This clarifies why (possibly weighted) estimation methods (e.g.
partial likelihood methods\cite{partial}) for contrasts in 
cause-specific hazards -- estimands relative to which competing events are 
\textsl{not} censoring events by definition \eqref{def: definition1} -- can be used to target contrasts in hazards under elimination of competing events -- estimands relative to which competing events \textsl{are} censoring
events by this definition (see Section \ref{censdef}).

We have now established that, under Figure \ref{dagimperfectfor14}, contrasts in all three
types of counterfactual hazards,  may be identified under the observed data structure in Section \ref{obsdata} with loss to follow-up. However, as in the case when competing events do not exist \citep{Hernanfail,hazards}, contrasts of
hazards when competing events exist do not generally have a causal
interpretation, even under the restrictive assumptions depicted in Figure %
\ref{dagimperfectfor14}.  Specifically, consider a counterfactual subdistribution hazard ratio as in \eqref{eqn:hazard2contrast}. This
contrast is conditional on previous survival from the event of interest, as represented in Figure \ref%
{daghr}, which is the same as Figure \ref{dagimperfectfor14} but with a square box around the node $Y_k$ to represent conditioning.  By graphical d-separation rules \citep{pearldag,swigs,causalbook}, this hazard ratio
then quantifies not only the causal paths between $A$ and $Y_{k+1}$, but also the
non-causal path $A\rightarrow Y_k\leftarrow U\rightarrow Y_{k+1}$ created by conditioning on the common effect $Y_k$
of treatment $A$ and the unmeasured variable $U$.   Similar graphical arguments can be used to illustrate that
the counterfactual hazard contrasts \eqref{eqn:hazard1contrast}
or \eqref{eqn:hazard3contrast}, which includes cause-specific hazard ratios, will not generally have a causal interpretation
under the realistic assumption that there exist unmeasured common causes of
early and late status on the event of interest.  Note that even identification of the direct effect allows the presence of such common causes; as we discussed in Section \ref{casei}, Figure \ref{dagimperfectfor14} is consistent with identification of the direct effect \eqref{eqn:RD1}.

\begin{figure}[tbp]
\centering
\includegraphics[scale=1]{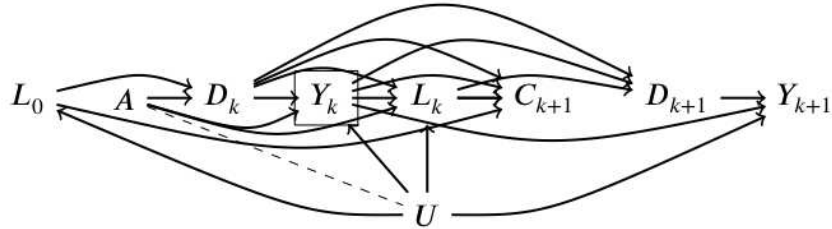}
\caption{Graphical illustration of why contrasts in counterfactual hazard ratios may not have a causal interpretation even under conditions that give identification of contrasts in any estimand in Table 1 under different levels of $a$ and $\overline{c}=0$.   }  
\label{daghr}
\end{figure}

\section{Choosing a definition of causal effect when competing events exist}\label{choosing}
We have explicitly defined two possible definitions of a causal treatment effect on the event of interest when competing events exist: the total effect and a direct effect.  Both are defined as contrasts in counterfactual risks.  In this section, we consider the choice between these two types of causal effects.  

As discussed in Section \ref{identification}, the total effect of treatment on the risk of the event of interest \eqref{eqn:RD2} can be identified under weaker untestable assumptions than those required for identification of the direct effect \eqref{eqn:RD1}.  Further, as discussed in Section \ref{caseii}, identifying assumptions for the total effect are expected to hold by design in a trial where $A$ is randomly assigned and there is no loss to follow-up.   However, the total effect may have an uncertain interpretation when the treatment affects the competing event.

To see this, consider the following extreme case.  Suppose that all individuals in a study population of patients with prostate cancer would die immediately from cardiovascular disease if they were assigned estrogen therapy (i.e. under $a=1$).  That is, for all individuals in the study population $D_1^{a=1,\overline{c}=\overline{0}}=1$.  As it is impossible to die of prostate cancer after dying from cardiovascular disease, none of these individuals would die of prostate cancer (have the event of interest) under $a=1$; that is, $\Pr [Y_{k+1}^{a=1,\overline{c}=\overline{0}}=1]=0$ for all $k$.  Suppose further that no individual would die from any cause other than prostate cancer during the study period if they were assigned placebo (i.e. under $a=0$).  That is, for all individuals, $\overline{D}_{K+1}^{a=0,\overline{c}=\overline{0}}=0$.  If at least one individual would die of prostate cancer by $k+1$ then $0<\Pr [Y_{k+1}^{a=0,\overline{c}=\overline{0}}=1]\leq1$.  Then the risk difference \eqref{eqn:RD2} will be negative and the total effect of assignment to estrogen therapy versus placebo on risk of prostate cancer death is protective.  

However, the total effect does not help us understand the reason for this protection; in this extreme case, the protection occurs only because estrogen therapy instantly kills everyone due to cardiovascular disease (such that they cannot die of prostate cancer).  Reporting both the total effect of treatment on the event of interest \eqref{eqn:RD2} and the total effect of treatment on the risk of the competing event \eqref{eqn:RD4} can alert to the existence of this problem.  In our example, \eqref{eqn:RD2} is negative but \eqref{eqn:RD4} is positive.  

By contrast, the direct effect \eqref{eqn:RD1} quantifies an effect of treatment on the event of interest that is not mediated by the competing event.  Therefore, a negative value of the direct effect cannot be explained (either wholly or in part) by a harmful effect of treatment on the competing event.  However, this direct effect is only well-defined -- and, therefore, only possible to confirm in a real-world study --
if there exist sufficiently well-defined interventions to eliminate losses
to follow-up \textsl{and} competing events. While we might imagine a study in which we
could eliminate loss to follow-up (e.g. by investing more financial
resources for follow-up), it is often difficult to imagine a study in which we
could eliminate competing events such as death.  

Other types of competing events may, in principle, be intervened on \citep{geskus}.  For example, in a study of the effect of a treatment on nosocomial infection during a given hospital admission, discharge from that admission is one competing event. However, while an intervention might prevent discharge in some patients, one that eliminates discharge in all patients (including those sufficiently recovered) is unrealistic.  When interventions $\overline{d}=0$ are not well-defined or are not realistically implemented, the plausibility of untestable assumptions required for identification may be especially questionable\citep{waterkill}.  For example, we noted in Section \ref{casei} concerns over the consistency assumption when $\overline{d}=0$ includes the \textsl{impossible} intervention elimination of death.  Still, positivity violations may occur when $\overline{d}=0$ includes the \textsl{unrealistic} intervention elimination of discharge because, in real-world studies, patients with $\overline{L}_k$ indicating recovery will have a zero (or close to zero) probability of not being discharged.   In Section \ref{Discussion} and Appendix \ref{crossworld}, we briefly consider alternative definitions of a treatment effect on the event of interest that does not capture the treatment's effect on the competing event.  Unlike any of the estimands in Table 1, these alternative definitions are based on contrasts in estimands defined by cross-world counterfactuals; their definition requires simultaneous knowledge of counterfactual outcomes for the same individual under different treatment interventions.  

In some cases, we may reasonably assume \textsl{a priori} that there is no effect of the treatment on the competing event; that is, we can remove arrows from treatment into future competing events on the causal DAG.   For example, this might be the case when interest is in the effect of a medical treatment on a disease outcome in an otherwise healthy population and the only deaths during the follow-up are due to car accidents.  In such cases, the total effect \eqref{eqn:RD2} does quantify the effect of the treatment on the event of interest that does not capture any treatment effect on the competing event.  Further, in some settings, mediation by competing events may not as materially complicate the interpretation of the total effect.  For example, knowing whether a treatment reduces the risk of nosocomial infection \textsl{only} by increasing the risk of hospital discharge might be less important (at least to a doctor or patient) than knowing whether a treatment reduces the risk of prostate cancer death \textsl{only} by increasing the risk of cardiovascular disease death. 

Finally, consider the effect of the treatment on the risk of the composite outcome \eqref{eqn:RD3}. Like the total effect \eqref{eqn:RD2}, this estimand does not require conceptualizing interventions on competing events and is identified by design in a trial where $A$ is randomized and loss to follow-up is absent.  Further, because the competing event is now redefined as part of the outcome, there is no longer the possibility of mediation by the competing event.   However, redefining effects through composite outcomes does not generally answer the question of interest; e.g. if the event of interest is prostate cancer death, \eqref{eqn:RD3} quantifies the effect of treatment on all-cause mortality and not the effect of treatment on prostate cancer death.  As noted in Section \ref{estimand}, in the case of mutually competing events, the effects \eqref{eqn:RD2} and \eqref{eqn:RD4} sum to \eqref{eqn:RD3}.  Thus, when these effects have opposite signs, \eqref{eqn:RD3} will be closer to zero than \eqref{eqn:RD2} or \eqref{eqn:RD4}.

\section{Estimation of direct and total effects}\label{stat}
As we established in Sections \ref{casei} and \ref{caseii}, the functions \eqref{eqn:gform1} and \eqref{eqn:gform2taub}, respectively, correspond to two
versions of the g-formula for risk had $A$ been set
to $a$ and censoring been eliminated.  The parametric g-formula\citep{robinsfail,taubman,documentation} and inverse probability weighting\citep{robinsfink,coxmsm} are two possible approaches to estimating the g-formula and associated contrasts.  Below, we briefly review some possible implementations of these approaches for the direct effect \eqref{eqn:RD1} and total effect \eqref{eqn:RD2}. 

\subsection{Direct effects}\label{directstat}
A parametric g-formula estimator of \eqref{eqn:gform1} directly estimates under model constraints
the components of the g-formula expression then uses Monte Carlo simulation based on the
estimated conditional densities of $L_k$ to approximate the high-dimensional sum/integral
over all risk factor histories \citep{taubman,pgformyoung,documentation}.  If we can assume exchangeability given only $L_0$ then \eqref{eqn:gform1} reduces to only a function of the observed cause-specific hazards of the event of interest and no Monte Carlo simulation is required.   Such a simplified parametric g-formula estimator of \eqref{eqn:gform1} is
\begin{equation}
\frac{1}{n}\sum_{i=1}^{n}\sum_{k=0}^{K}p(a,l_{0i},k;\hat{\theta})\prod_{j=0}^{k-1}\left[1-p(a,l_{0i},j;\hat{\theta})\right]\label{eqn:gcomp1} 
\end{equation}
where $p(a,l_{0i},k;\theta)$ is a model (e.g. a pooled over time logistic regression model\citep{dagostino}) indexed by parameter $\theta$ for the observed conditional cause-specific hazard of the event of interest $\Pr [Y_{k+1}=1|\overline{C}_{k+1}=\overline{D}_{k+1}=\overline{Y}%
_{k}=0,A=a,L_0]$ evaluated at treatment level $a$ and individual $i$'s observed values of the baseline covariates $l_{0i}$ with $\hat{\theta}$ a consistent estimator of the true $\theta$.  Alternative implementations of the parametric g-formula can be based on assumptions of continuous time hazard models; e.g. proportional hazards \citep{partial,breslow} or additive hazards \citep{aalen2008survival} models . 

The IPW representation \eqref{eqn:ipw1} of the g-formula \eqref{eqn:gform1}  helps to motivate an alternative approach to the parametric g-formula for estimating \eqref{eqn:gform1} and corresponding contrasts under different levels of $a$.  Following previous authors \citep{robrot92,robinsinfo,robinsfink}, an IPW estimator can be constructed as the complement of a weighted product-limit (Kaplan-Meier) estimator \citep{km} as follows, with $(a_i,\overline{l}_{Ki},\overline{c}_{K+1i},\overline{d}_{K+1i},\overline{y}_{K+1i})$ individual $i$'s values of $(A,\overline{L}_{K},\overline{C}_{K+1},\overline{D}_{K+1},\overline{Y}_{K+1})$:
\begin{equation}
\sum_{k=0}^{K}\hat{h}_{k}(a;\hat{\alpha},\hat{\eta})\prod_{j=0}^{k-1}\left[1-\hat{h}_{j}(a;\hat{\alpha},\hat{\eta})\right]  \label{eqn:ipw1est}
\end{equation}
where $\hat{h}_{k}(a;\hat{\alpha},\hat{\eta})=\frac{\sum_{i=1}^{n}y_{k+1i}(1-y_{ki})w_{ki}(\hat{\alpha},\hat{\eta})I(a_i=a)}{\sum_{i=1}^{n}(1-y_{ki})w_{ki}(\hat{\alpha},\hat{\eta})I(a_i=a)}$, $w_{ki}(\hat{\alpha},\hat{\eta})=\frac{\prod_{j=0}^k(1-c_{j+1i})(1-d_{j+1i})}{\prod_{j=0}^k[1-r(a,\overline{l}_{ji},j;\hat{\alpha})][1-q(a,\overline{l}_{ji},j;\hat{\eta})]}$, $r(a,\overline{l}_{ji},j;\alpha)$ and $q(a,\overline{l}_{ji},j;\eta)$ are models (e.g. pooled over time logistic regression models) indexed by parameters $\alpha$ and $\eta$ for the observed cause-specific loss to follow-up $\Pr[C_{j+1}=1|%
\overline{L}_j=\overline{l}_{ji},\overline{C}_{j}=\overline{D}_{j}=\overline{Y}%
_{j}=0,A=a]$ and competing event $\Pr[D_{j+1}=1|%
\overline{L}_j=\overline{l}_{ji},\overline{C}_{j+1}=\overline{D}_{j}=\overline{Y%
}_{k}=0,A=a]$ hazards with $\hat{\alpha}$ and $\hat{\eta}$ consistent estimators of $\alpha$ and $\eta$, respectively. 

Unlike the parametric g-formula estimator, which relies on correctly specified models for time-varying observed cause-specific event of interest hazards, as well as (generally) the joint conditional distributions of the time-varying confounders $L_k$, consistency -- here, meaning statistical convergence rather than \eqref{eqn:consistency1} or \eqref{eqn:consistency2} -- of the IPW estimator requires only correct specification of the weight denominator models (for the observed loss to follow-up and competing event hazards).  An alternative approach based on a stabilized weighting scheme \citep{coxmsm} that multiplies the numerator of $w_{ki}(\hat{\alpha},\hat{\eta})$ by any function of, at most, $a$ and $k$ may reduce variability of the weights. Note that, unlike the parametric g-formula, the algorithmic complexity of IPW estimation is not substantially reduced under adjustment for only $L_0$.  In this case, we simply replace functions of $\overline{L}_j$ with functions of $L_0$ in the weight denominator models.   Alternative estimators of \eqref{eqn:gform1} and corresponding contrasts under different levels of $a$ with double-robust properties follow from previous work by treating competing events like loss to follow-up \citep{robrot92,robinsinfo,unified,Bangandrobins,schnitzertmle}.  Extensions that allow construction of sensitivity bounds under violations of exchangeability \eqref{eqn:C1} have also been derived \citep{scharfstein99}. 

\subsection{Total effects}\label{totalstat}

A key distinction between expressions \eqref{eqn:gform1} and \eqref{eqn:gform2taub} is that the latter depends on knowledge of the time-varying observed cause-specific hazards of the competing event $\Pr[%
D_{k+1}=1|\overline{L}_k=\overline{l}_{k},\overline{D}_{k}=\overline{C}_{k+1}=\overline{Y}%
_{k}=0,A=a]$ while the
former does not. Thus, a parametric g-formula estimator for the latter will
rely on an estimate of this quantity -- e.g. correct specification of the model $q(a,\overline{l}_{j},j;\eta)$ as defined in the previous section -- while the former will not. See \cite{documentation} and \cite{gformrpackage} for implementations of the parametric g-formula in SAS and R, respectively, for estimating either
\eqref{eqn:gform1} or \eqref{eqn:gform2taub} and associated contrasts (risk differences/risk ratios).   

Similarly, the algorithmic complexity of this approach simplifies considerably when exchangeability given only $L_0$ is assumed.   In this restricted case, expression \eqref{eqn:gform2taub} will only depend on the time-varying observed cause-specific hazards of the event of interest and the competing event.  A parametric g-formula estimator in this case is
\begin{equation}
\frac{1}{n}\sum_{i=1}^{n}\sum_{k=0}^{K}p(a,l_{0i},k;\hat{\theta})\left[1-q(a,l_{0i},k;\hat{\eta})\right]\prod_{j=0}^{k-1}\left[1-p(a,l_{0i},j;\hat{\theta})\right] \left[1-q(a,l_{0i},j;\hat{\eta})\right]\label{eqn:gcomp2} 
\end{equation}
where, as above, $p_{k}(a,l_{0},k;\theta)$ and $q_{k}(a,l_{0},k;\eta)$ are models for the observed event of interest and competing event cause-specific hazards, respectively conditional on $A=a$, $L_0=l_0$ and remaining not lost.

We considered two different algebraically equivalent IPW representations
of the g-formula \eqref{eqn:gform2taub}, motivating two different IPW approaches.  Following the algebraic equivalence between the g-formula \eqref{eqn:gform2taub} and the IPW expression \eqref{eqn:ipw2first}, an IPW estimator based on weighted estimates of the observed subdistribution hazards of the event of interest is:
\begin{equation}
\sum_{k=0}^{K}\hat{h}^{sub}_{k}(a;\hat{r}^{sub})\prod_{j=0}^{k-1}\left[1-\hat{h}^{sub}_{j}(a;\hat{r}^{sub})\right]
\label{eqn:ipw2firstest}
\end{equation}
where $\hat{h}^{sub}_{k}(a;\hat{r}^{sub})=\frac{\sum_{i=1}^{n}y_{k+1i}(1-y_{ki})w^{sub}_{ki}(\hat{r}^{sub})I(a_i=a)}{\sum_{i=1}^{n}(1-y_{ki})w^{sub}_{ki}(\hat{r}^{sub})I(a_i=a)}$, $w^{sub}_{ki}(\hat{r}^{sub})=\frac{\prod_{j=0}^k(1-c_{j+1i})}{\prod_{j=0}^k[1-\hat{r}^{sub}(a,\overline{l}_{ji},\overline{d}_{ji},j)]}$, and $\hat{r}^{sub}(a,\overline{l}_{ji},\overline{d}_{ji},j)=r(a,\overline{l}_{ji},j;\hat{\alpha})$ (as above, a model-based estimate of the loss to follow-up hazard $\Pr[C_{j+1}=1|\overline{L}_j=\overline{l}_{ji},\overline{C}_{j}=\overline{D}_{j}=\overline{Y}_{j}=0,A=a]$) when $\overline{d}_{ji}=0$ and, otherwise,  $\hat{r}^{sub}(a,\overline{l}_{ji},\overline{d}_{ji},j)$ is an estimate of $\Pr[C_{j+1}=1|\overline{L}_j=\overline{l}_{ji},\overline{C}_{j}=\overline{Y}_{j}=0,\overline{D}_j,A=a]$ amongst those with $\underline{D}_s=1$, for some $s\leq j$.  The latter must always take the value zero because once an individual experiences the competing event, he cannot be subsequently lost (see Section \ref{censdef}).   

Following the algebraic equivalence of the g-formula \eqref{eqn:gform2taub} and \eqref{eqn:ipw2}, an alternative IPW estimator based on weighted estimates of the observed cause-specific hazards of the event of interest and the competing event is:
\begin{equation}
\sum_{k=0}^{K}\hat{h}^{1}
_{k}(a;\hat{\alpha})\{1-\hat{h}^{2}_{k}(a;\hat{\alpha})\}\prod_{j=0}^{k-1}[\{1-\hat{h}^{1}_{j}(a;\hat{\alpha})\}\{1-\hat{h}^{2}_{j}(a;\hat{\alpha})\}]
\label{eqn:ipw2est}
\end{equation}%
where $\hat{h}^{1}_{k}(a;\hat{\alpha})=\frac{\sum_{i=1}^{n}y_{k+1i}(1-y_{ki})(1-d_{ki})w^{c}_{ki}(\hat{\alpha})I(a_i=a)}{\sum_{i=1}^{n}(1-y_{ki})(1-d_{ki})w^{c}_{ki}(\hat{\alpha})I(a_i=a)}$, $\hat{h}^{2}_{k}(a;\hat{\alpha})=\frac{\sum_{i=1}^{n}d_{k+1i}(1-y_{ki})(1-d_{ki})w^{c}_{ki}(\hat{\alpha})I(a_i=a)}{\sum_{i=1}^{n}(1-y_{ki})(1-d_{ki})w^{c}_{ki}(\hat{\alpha})I(a_i=a)}$, $w^{c}_{ki}(\hat{\alpha})=\frac{\prod_{j=0}^k(1-c_{j+1i})}{\prod_{j=0}^k[1-r(a,\overline{l}_{ji},j;\hat{\alpha})]}$ with, as above, $r(a,\overline{l}_{j},j;\alpha)$ a model for the  loss to follow-up hazard.  The estimator \eqref{eqn:ipw2est} coincides with a weighted Aalen-Johansen estimator \citep{aalenjohansen}.  Analogously, consistency of either the IPW estimator \eqref{eqn:ipw2firstest} or \eqref{eqn:ipw2est} of the g-formula \eqref{eqn:gform2taub} requires fewer model assumptions than those required for the corresponding parametric g-formula estimator.  Also, as above, the general algorithmic complexity of either of the above IPW estimators does not substantially change under adjustment for only $L_0$; we simply replace functions of $\overline{L}_j$ with functions of $L_0$ in the weight denominator models.   

Various estimators of contrasts in \eqref{eqn:gform2taub} for different treatment interventions given by \cite{Bekaertcr}, \cite{lokcif}, \cite{Moodiecr} and Appendix D of \cite{repintyoung} allow exchangeability to depend on time-varying $\overline{L}_k$.  Alternative estimators of \eqref{eqn:gform2taub} under exchangeability given only $L_0$ follow from results in \cite{finegray}.   See Appendix \ref{dataappendix} for a discussion of parametric g-formula or IPW estimators of the risk of the competing event itself \eqref{eqn:risk3} and corresponding effects \eqref{eqn:RD4}.

\section{Example: A randomized trial on prostate cancer therapy}\label{analysis}

We illustrate the ideas outlined above using data
from a trial that randomly assigned estrogen therapy, Diethylstilbestrol (DES), or placebo to prostate cancer patients.  These data are
available at \url{http://biostat.mc.vanderbilt.edu/DataSets} and have been
used in other methodological articles on competing risks \citep{prostatedata,fine1999analysing,kay1986treatment,harrell1996multivariable}. In this trial,
502 patients were assigned to four different treatment arms.  Here, we restrict our analysis to the 125 patients in the high-dose estrogen therapy arm ($A=1$) and the 127 patients in the placebo arm ($A=0$).  Interest is in whether treatment has a causal effect on prostate cancer death.  Death due to other causes is therefore a competing event. Follow up data on event status was available in monthly intervals. We considered effects through follow-up month $K+1=60$ (5-years post-randomization).  In the treatment arm, 26 patients died of prostate cancer and 63 died of other causes.  In the placebo arm, 35 patients died of prostate cancer while 54 died of other causes.    

As in many randomized trials, only baseline covariates ($L_0$) were available, which implies that our exchangeability assumptions need to hold conditional on baseline covariates only. The variables included in $L_0$ were daily activity function, age, hemoglobin level, and history of cardiovascular disease. For simplicity, in this illustrative example, we categorized continuous covariates to reduce the dimension of $L_0$ (limiting the possible model functional forms but at the expense of a less reasonable version of exchangeability, particularly for the direct effect).

We applied the parametric g-formula and IPW estimators described in Section \ref{stat} to estimate total and direct treatment effects of estrogen therapy on prostate cancer death based on pooled over time logistic models $p_{k}(a,l_{0},k;\theta)$ for the observed cause-specific hazard of the event of interest, $q_{k}(a,l_{0},k;\eta)$ for the observed cause-specific hazard of the competing event and $r(a,l_{0},k;\alpha)$ for the observed cause-specific hazard of loss to follow-up.  Additional implementation details for all estimators, including structure of input data sets and model assumptions, are given in Appendix \ref{dataappendix} along with a description of the parametric g-formula and IPW estimators of the risk of the competing event itself and the corresponding total treatment effects. R code for all analyses is provided in the supplementary materials.

\subsection{Total effect of estrogen therapy on prostate cancer death}
We estimated the risk of prostate cancer death without elimination of competing events \eqref{eqn:risk2} under assignment to estrogen therapy ($a=1$) and placebo ($a=0$) for each follow-up month $k+1=1,...60$ using a parametric g-formula estimator as defined in \eqref{eqn:gcomp2}, an IPW estimator as defined in \eqref{eqn:ipw2firstest} and an alternative IPW estimator as defined in \eqref{eqn:ipw2est}.  Total effects were estimated as contrasts (ratios and differences) in these estimated risks across treatment levels.

Because no loss to follow-up occurred before month 50, the risk without elimination of competing events \eqref{eqn:risk2} up to month 50 is identified by the observed data function $\Pr[Y_{k+1}=1|A=a], k<50$, which is consistently estimated by simply the number of those with the event of interest by $k+1$ divided by the number of individuals randomized to $A=a$. Both of our IPW estimators incorporate the knowledge that the hazard of censoring was zero before month 50 (see Appendix \ref{dataappendix}) so the estimates are guaranteed to be identical to each other (and the nonparametric cumulative proportion estimator) before month 50.   
  
\begin{figure}[h!]
\centering
\includegraphics[scale=.6]{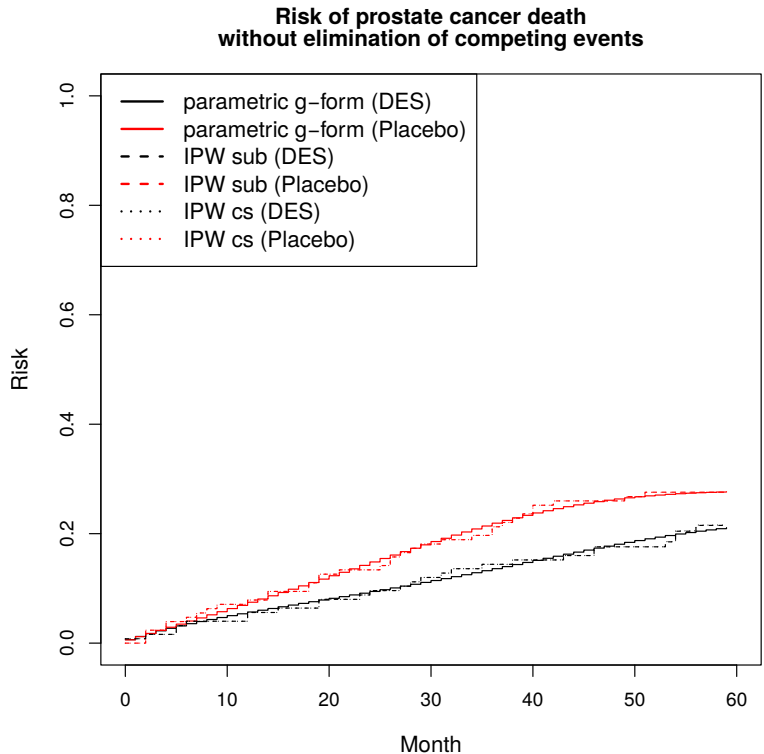}
\caption{Parametric g-formula estimates \eqref{eqn:gcomp2}, IPW estimates \eqref{eqn:ipw2firstest} (\textsl{IPW sub}) and IPW estimates \eqref{eqn:ipw2est} (\textsl{IPWcs}) of the the risk of prostate cancer death without elimination of competing events by follow-up month $k+1=1,\ldots60$ under high-dose estrogen therapy (DES) versus placebo. }
\label{cemevent}
\end{figure}

Total effect estimates by 60 month follow-up, calculated as contrasts in estimates of the risk $\Pr [Y_{60}^{a,\overline{c}=\overline{0}}=1] $, are presented in Table \ref{results}. All 95\% confidence intervals were calculated from the 2.5 and 97.5 percentiles of a nonparametric bootstrap distribution based on 500 bootstrap samples. While 95\% confidence intervals are wide, all point estimates are in line with a total protective effect of estrogen therapy compared with placebo on the risk of prostate cancer death.   Figure \ref{cemevent} shows that the estimated risks under treatment and placebo start to diverge shortly after the start of the follow-up. 

\begin{table}[!htp]
\caption{Parametric g-formula and inverse probability weighted (IPW) estimates of the total treatment effect on the risk of the event of interest, total treatment effect on the risk of the competing event and the direct effect on the risk of the event of interest by 60-month follow-up. \protect\label{results}}
\label{wtdist}
\begin{center}
\begin{tabular}{lcc}
\hline
\noalign{\smallskip}  & Risk ratio (95\% CI$^{*}$) & Risk difference (95\% CI$^{*}$)    \\ \hline
\textbf{Total effect (prostate cancer death)} &  & \\ 
parametric g-formula & $0.76$ $(0.47, 1.24)$ & $-0.07$ $(-0.18, 0.05) $\\ 
IPW cs$^{**}$ & $0.78$ $(0.49, 1.28) $ & $ -0.06$ $(-0.17, 0.06) $\\
IPW sub$^{***}$ & $0.78$ $(0.47, 1.32)$ & $ -0.06$ $(-0.18, 0.06) $\\  
\textbf{Total effect (other causes of death)} &  & \\ 
parametric g-formula & $1.28$ $(0.97, 1.61)$ & $0.12$ $(-0.01, 0.23)$\\ 
IPW cs$^{**}$ & $1.19$ $(0.91, 1.54) $ & $0.08$ $(-0.04, 0.20) $ \\
IPW sub$^{***}$ & $1.19$ $(0.90, 1.54)$& $0.08$ $(-0.05, 0.21)$ \\  
\textbf{Direct effect (prostate cancer death)} &  & \\ 
parametric g-formula & $0.91$ $(0.57, 1.47)$ & $-0.03$ $(-0.19, 0.14)$ \\ 
IPW  & $0.98$ $(0.56, 1.59)$ & $-0.01$ $(-0.20, 0.17) $\\   \hline
\end{tabular}%
\end{center}
\par
{\footnotesize {$^{*}$Percentile-based bootstrap from 500 bootstrap samples
}}
\par
{\footnotesize {$^{**}$Based on estimator \eqref{eqn:ipw2est} for prostate cancer death and \eqref{eqn:ipw2estce} for other death
}}
\par
\par
{\footnotesize {$^{***}$Based on estimator \eqref{eqn:ipw2firstest} for prostate cancer death and the same estimator for other death with other death treated as event of interest
}}
\par
\end{table}

Because some patients in both arms died of other causes, our estimates cannot rule out that the total effect may be partly explained by a harmful effect of estrogen therapy on death due to other causes (this would be true even if all of our assumptions held and sampling variability could not explain our findings). Therefore, we also estimated the total effect of estrogen therapy versus placebo on the risk of other causes of death as in \eqref{eqn:RD4}.  We used the parametric g-formula and two versions of IPW to estimate the risk of the competing event itself under treatment and placebo\eqref{eqn:risk3}, relying on the same pooled over time logistic regression models for the observed hazards of prostate cancer death, other death and loss to follow-up referenced above  (see Appendix \ref{dataappendix}).

\begin{figure}[h!]
\centering
\includegraphics[scale=.6]{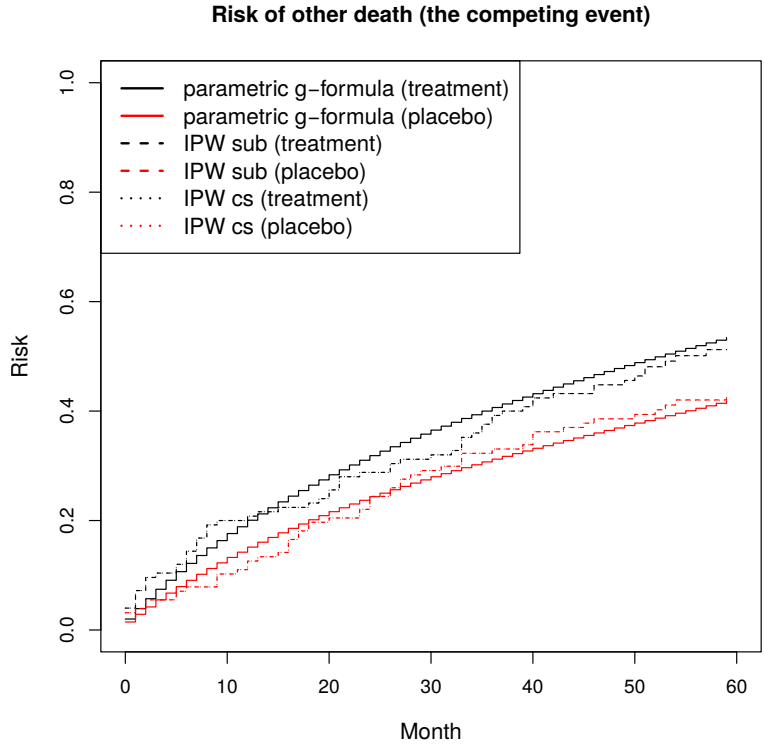}
\caption{Parametric g-formula estimates \eqref{eqn:gcomp2ce}, IPW estimates \eqref{eqn:ipw2estce} (\textsl{IPWcs}) and IPW estimates of \eqref{eqn:ipw2firstest} but replacing the event of interest with the competing event (\textsl{IPW sub}) of the risk of other causes of death by follow-up month $k+1=1,\ldots60$ under high-dose estrogen therapy (DES) versus placebo. }
\label{riskce}
\end{figure}

Estimates of the total effect of treatment on other causes of death by 60 months and 95\% confidence intervals are shown in Table \ref{results}.    Again, while all 95\% confidence intervals are wide, point estimates are in line with a harmful total effect of treatment on the competing event.  Figure \ref{riskce} shows that the estimated risks under treatment versus placebo diverge from the start of the follow-up. Not surprisingly, the curves based on the parametric g-formula, which rely on pooled over time models for the event of interest and competing event hazards, are smoother. 

\begin{figure}[h!]
\centering
\includegraphics[scale=.6]{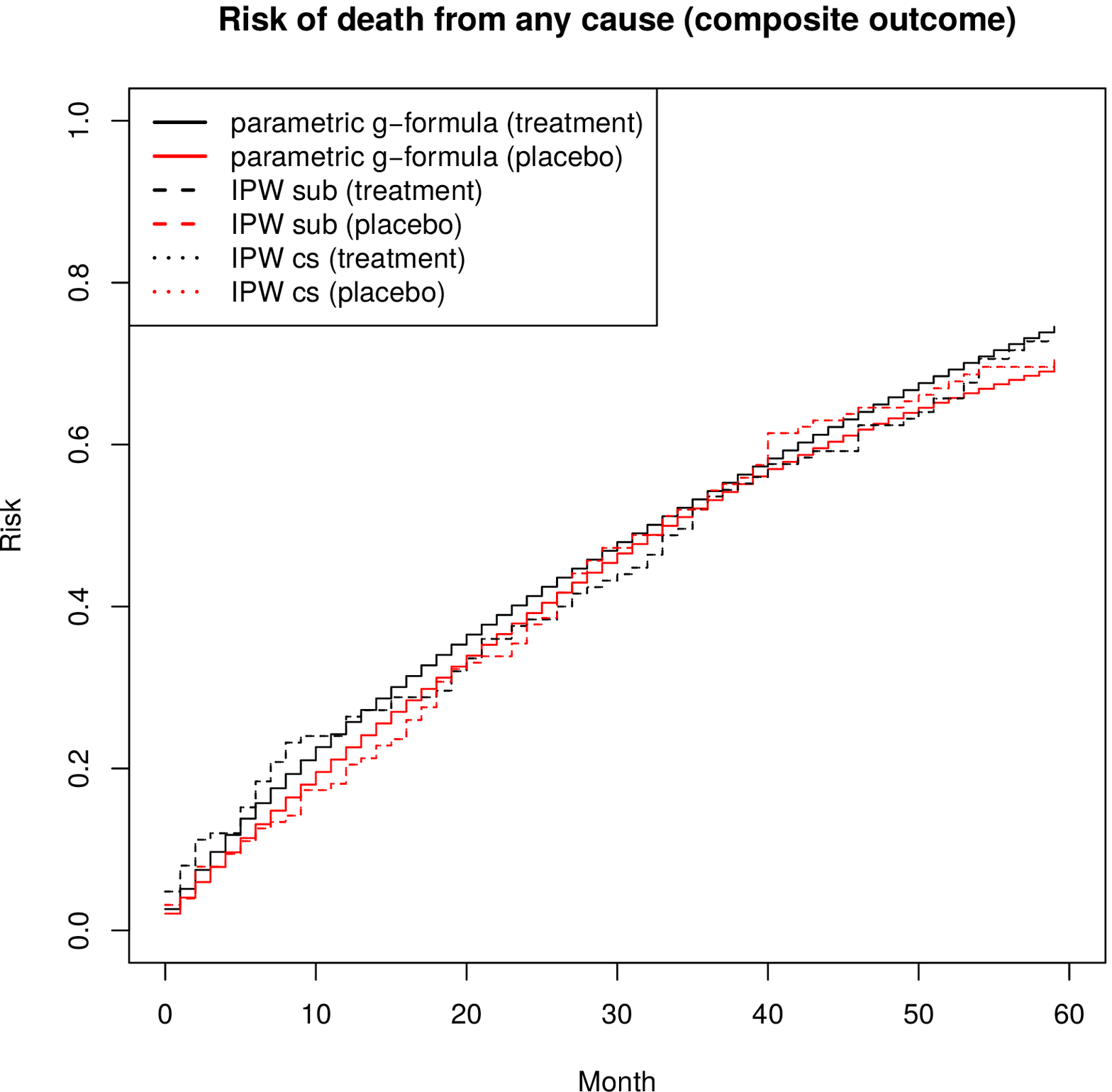}
\caption{Parametric g-formula and IPW estimates of the risk of the composite outcome (based on sums of corresponding estimates of the risk of prostate cancer death without elimination of competing events and risk of other causes of death) by follow-up month $k+1=1,\ldots60$ under high-dose estrogen therapy (DES) versus placebo. }
\label{riskcomp}
\end{figure}

These estimates suggest that our estimates of a protective total effect on prostate cancer death may be attributable to an effect of treatment on other causes of death. In fact, estimates of the effect on the composite outcome death from any cause are close to the null through most of the follow-up (as shown in Figure \ref{riskcomp}) due to the opposite directions of the total effect estimates on prostate cancer death and other causes of death.

\subsection{Direct effect of estrogen therapy on prostate cancer death}
We estimated the direct effect (as in \eqref{eqn:RD1} on the additive scale) of estrogen therapy on prostate cancer death not mediated by the effect of treatment on other causes of death such that competing events are censoring events. We estimated this effect via the parametric g-formula estimator \eqref{eqn:gcomp1} and the IPW estimator \eqref{eqn:ipw1est} using the same pooled over time logistic regression model assumptions as in the previous section.

\begin{figure}[h!]
\centering
\includegraphics[scale=.6]{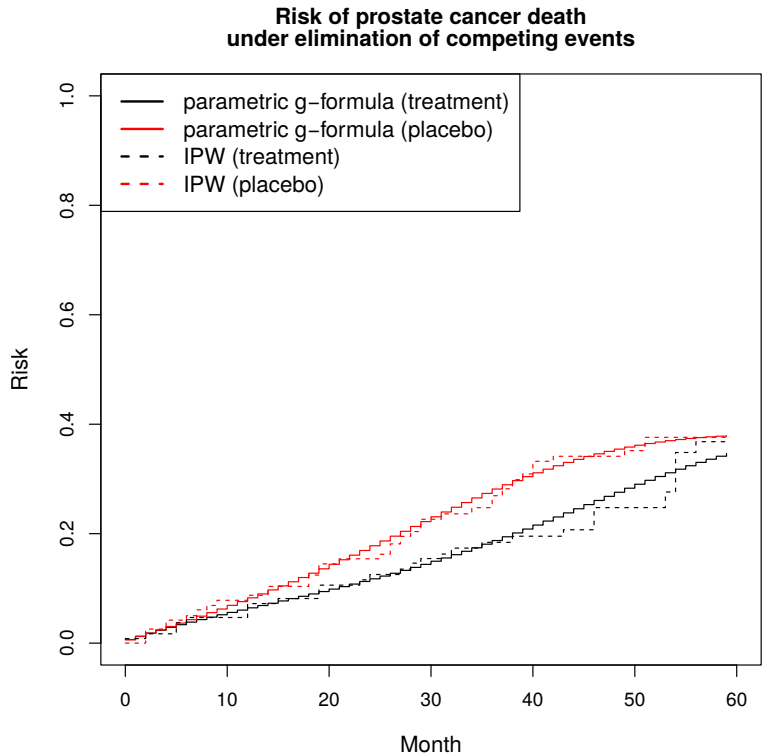}
\caption{Parametric g-formula estimates \eqref{eqn:gcomp1} and IPW estimates \eqref{eqn:ipw1est} of the risk of prostate cancer death under elimination of competing events by follow-up month $k+1=1,\ldots60$ under high-dose estrogen therapy (DES) versus placebo. }
\label{ceievent}
\end{figure}

The direct effect estimates, shown in Table \ref{results}, suggest the effect of treatment on the risk of prostate cancer death under a hypothetical scenario in which the competing event (other causes of death) is removed is smaller by 60-month follow-up than the total effect on prostate cancer death. As shown in Figure \ref{ceievent}, there is some variation over time.   

However, these estimates need to be interpreted with caution, not only because the 95\% confidence intervals are wide and the possibility of model misspecification, but because they rely on particularly strong identifying assumptions (those of Section \ref{casei}) and they cannot be confirmed in any real-world study.  As we discussed in Section \ref{choosing}, elimination of the competing event is impossible, which makes these effect estimates ill-defined and violations of required identifying assumptions likely.  Here, the assumption that only covariates included in $L_0$ are sufficient to ensure exchangeability for censoring by death is particularly strong. Bias might have been mitigated by control for post-randomization (time-varying) prognostic factors for prostate cancer death that also affect death from other causes had they been measured in this study.   

\section{Discussion}\label{Discussion}

In this paper, we used a counterfactual framework for causal inference to formally define classical statistical estimands from the competing risks literature.  When competing events are defined as censoring events, a contrast in counterfactual risks of the event of interest quantifies a direct effect of treatment on the event of interest that does not capture the treatment's effect on the competing event.  Otherwise, it quantifies a total effect of treatment on this event that captures all causal pathways between treatment and the event of interest.  

We formalized conditions under which these total and direct effects may be identified.  In particular, when the estimand is defined as a total effect and there is loss to follow-up in the study, competing events are time-varying covariates which are needed to ensure exchangeability, together with other baseline and time-varying covariates, even in randomized trials.  For both types of effects, we showed that identifying functions coincide with different versions of Robins's g-formula [\citeyear{robinsfail}].  We linked these identification results to various possible estimators of total and direct effects.  Previous authors have argued that ``...there are no established rules for representing competing risks on a DAG.'' \citep{leskolau}.   We showed how exchangeability assumptions in this setting can be assessed on a causal diagram using established graphical rules for identification once the estimand is made explicit \citep{pearldag,swigs}.  

We presented an application of some of these estimators to data from a trial of estrogen therapy and prostate cancer death.  As in many randomized trials, only baseline covariates were available, which makes exchangeability assumptions less plausible \citep{ppenejm,ppemiguelsonia}. We used an example of a randomized trial to motivate ideas but our results trivially extend to observational studies in which treatment assignment is not randomized by the investigator.

While this paper highlighted the central role of the scientific question to determine whether a competing event is a censoring event, this choice does not uniquely apply to competing events.  Many other events observed during the course of a longitudinal study may involve such a choice.  For example, here we considered counterfactual outcomes indexed by interventions only on baseline assignment to a given treatment strategy such that all effects considered are \textsl{intention to treat effects}.  Had we considered counterfactual outcomes indexed by elimination of nonadherence, which can be used to define \textsl{per-protocol effects} \citep{ppenejm,ppemiguelsonia}, then failure to adhere to the study protocol would have been a censoring event by the definition \eqref{def: definition1}.  

We also considered three definitions of the hazard of the event of interest at a given time under a counterfactual intervention on treatment.  Two of these hazard definitions coincide with the popular \textsl{cause-specific} and \textsl{subdistribution} hazards of the classic statistical literature.   At odds with previous recommendations advocating for the reporting of cause-specific hazard ratios when interest is in ``etiology'' \citep{latouche,austincirc,laucr}, we clarified that contrasts in any of these counterfactual hazards do not generally have a causal interpretation, even when assumptions that give identification of these contrasts hold in the study.  This failure of causal interpretation occurs even when the restrictive assumptions that identify a direct effect on the risk of the event of interest hold.  In line with previous recommendations in the non-competing risk setting \citep{Hernanfail,hazards}, we do not recommend the routine use of any hazard as a measure of causal effect when competing events exist due to these problems.  However, as we reviewed, many estimators of identifying functions for the total effect or direct effect for data with censoring events involve estimating different types of observed hazards.  In the case of total effects, the choice to use methods that estimate (possibly weighted or conditional) observed cause-specific or subdistribution hazards may impact model misspecification bias and precision but the target causal estimand remains a total effect.

We clarified that a competing event generally acts as a mediator of the treatment effect on the event of interest, leading to uncertain interpretation of the total effect.  Therefore, except in special cases where we are confident that there is no effect of the treatment on the competing event, effects that quantify a treatment effect not capturing the treatment's effect on the competing event are attractive.  We focused on one such effect, a \textsl{controlled} direct effect, that coincides with a contrast in estimands historically considered in the statistical literature.  As we discussed, the direct effect \eqref{eqn:RD1} relies on implausible or unrealistic interventions on competing events and, in turn, requires strong assumptions for identification in a real-world study.  Other effect definitions that can quantify an effect of the treatment not capturing the treatment's effect on the competing event have been advocated in the causal inference literature; in particular, the \textsl{survivor average causal effect} (SACE)\citep{rubinsace} which quantifies the average causal effect of the treatment on the event of interest among the subset of individuals in the study population who would never experience the competing event under any level of treatment. While such an effect does not require conceptualizing interventions on competing events, it quantifies a treatment effect in an unknown subset of the study population because it is only possible to observe an individual's competing event status under a single level of treatment (see Appendix \ref{crossworld}).

In conclusion, a counterfactual framework for causal inference elucidates choices in the analysis of failure time data with competing events.  Such a framework allows explicit definition and interpretation of causal treatment effects, consideration of the nature and strength of assumptions needed to identify these effects in real-world studies and, therefore, guidance for the selection of statistical methods.   Current options for defining a treatment effect on the event of interest that does not capture the treatment's effect on the competing event have significant drawbacks, either requiring ill-defined interventions in the study population or well-defined interventions in an unknowable subpopulation.  New effect definitions for evaluating treatment biological harm or benefit on the event of interest that avoid these problems will be an essential component of future work. 

\section*{Acknowledgements}
The authors thank Susan Gruber for helpful comments on an early version of the manuscript, as well as L. Paloma Rojas Saunero and Thomas A. Gerds for helpful literature references.  This work was funded by NIH grant R37 AI102634 and the Research Council of Norway, grant NFR239956/F20.

\bibliographystyle{plainnat}
\bibliography{refs}

\pagebreak

\appendix

\section{Single-world intervention graphs }\label{appasub}
\cite{swigs} defined a graphical condition based on a d-separation \citep{pearldag} relation that gives general identification of a time-varying treatment effect by a particular g-formula \citep{robinsfail}. They further show that, given an appropriate consistency assumption, this graphical condition for identification implies an exchangeability condition such as those in \eqref{eqn:C1} or \eqref{eqn:C2} of the main text under the assumption that the causal DAG represents an underlying counterfactual causal model \citep{robins2010alternative}.  Their d-separation condition is applied to a transformation of a causal DAG \citep{pearldag} representing assumptions on the underlying data generating process that produced the data in the observational study such as those in Figures \ref{dagimperfectfor14} and \ref{dag3} of the main text. \cite{swigs} call this transformation a Single World Intervention Graph (SWIG). We now illustrate how to evaluate identification of any of the counterfactual estimands in Table 1 of the main text using SWIGs.

To evaluate identification of any of the estimands in Table 1 of the main text, the following transformation is applied to the causal DAG representing the assumed underlying observed data generating mechanism:
\begin{enumerate}
\item Split each intervention node into two nodes with one node containing the \textsl{natural value} of the intervention variable (the value that would have been observed at $k$ were the intervention discontinued right before $k$ \citep{WHOchap,swigs,youngthreshold} ) and the other a fixed value under intervention.  Intervention variables in our case are treatment $A$ and censoring at each time (with censoring nodes including loss to follow-up and competing events or only loss to follow-up, depending on the estimand, as explained in the main text).  
\item Index all random variables after $A$ as counterfactuals under the intervention, including the natural values of the intervention variables.
\item All arrows originally out of the observed values of intervention variables on the original causal DAG should now be out of the intervention value and all arrows into the observed values on the original causal DAG should now be into the (counterfactual) natural value of that variable.
\end{enumerate}

\begin{figure}[tbp]
\centering
\includegraphics[scale=1]{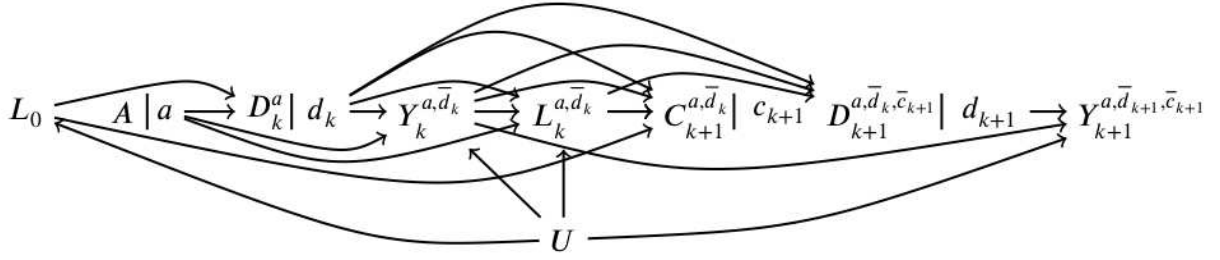}
\caption{A SWIG template $\mathcal{G}(a,\overline{c},\overline{d})$ derived from the causal DAG in Figure \ref{dagimperfectfor14} which implies exchangeability \eqref{eqn:C1} holds.  }
\label{swig1}
\end{figure}

Consider an intervention that sets $A$ to $a$ and somehow eliminates loss to follow-up and competing events.  Figure \ref{swig1} is a SWIG template \citep{swigs}, denoted $\mathcal{G}(a,\overline{c},\overline{d})$, which is a transformation of the causal DAG in Figure \ref{dagimperfectfor14}.  $\mathcal{G}(a,\overline{c},\overline{d})$ is a template because it corresponds to any intervention where $\overline{C}_{k+1},\overline{D}_{k+1}$ are set to fixed values $\overline{c}_{K+1},\overline{d}_{K+1}\equiv\overline{c},\overline{d}$.  We are interested specifically in $\overline{c}=\overline{d}=0$ (we consider the general template to reduce clutter on the graph) .  

\begin{figure}[tbp]
\centering
\includegraphics[scale=1]{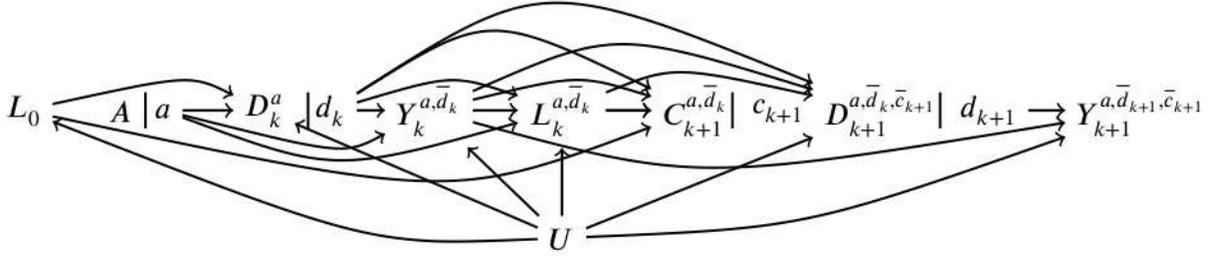}
\caption{A SWIG template $\mathcal{G}(a,\overline{c},\overline{d})$ derived from the causal DAG in Figure \ref{dag3} which implies exchangeability \eqref{eqn:C1} fails.  }
\label{swig2}
\end{figure}

Define $A^{a}$ as the value of treatment any individual would receive under an intervention that sets $A=a$ such that, by definition, $A^{a}\equiv a$ for all individuals. Also define $C_j^{a,\overline{c}_j,\overline{d}_{j-1}}$ and $D_j^{a,\overline{c}_j,\overline{d}_j}$ as the values of the censoring indicators at $j$ under an intervention that sets $(\overline{C}_{K+1},\overline{D}_{K+1})=(\overline{c}_{K+1},\overline{d}_{K+1})$ such that, by definition, $C_j^{a,\overline{c}_j,\overline{d}_{j-1}}\equiv c_j$ and $D_j^{a,\overline{c}_j,\overline{d}_j}\equiv d_j$ for all individuals.  Then, given the consistency assumption \eqref{eqn:consistency1} and for $\overline{c}=\overline{d}=0$, exchangeability \eqref{eqn:C1} is implied under the data generating assumptions of the causal DAG in Figure \ref{dagimperfectfor14} by the absence of any unblocked backdoor paths on the SWIG template $\mathcal{G}(a,\overline{c},\overline{d})$ (Figure \ref{swig1}) between:
\begin{enumerate}
\item the natural value of treatment $A$ and future counterfactual outcomes $\overline{Y}^{a,\overline{d},\overline{c}}_{K+1}$ conditional on $L_0$
\item for any $j=0,\ldots,K$, the natural value of the loss to follow-up indicator $C_{j+1}^{a,\overline{c}_{j},\overline{d}_{j}}$ and future counterfactual outcomes $\underline{Y}^{a,\overline{d}_{j+1},\overline{c}_{j+1}}_{j+1}$ conditional on $(\overline{L}^{a,\overline{c}_{j}=\overline{d}_{j}}_{j},\overline{Y}^{a,\overline{c}_{j},\overline{d}_{j}}_{j},\overline{D}_j^{a,\overline{c}_j,\overline{d}_j}\equiv d_j,\overline{D}_j^{a,\overline{c}_j,\overline{d}_{j-1}}, \overline{C}_j^{a,\overline{c}_j,\overline{d}_{j-1}}\equiv c_j, \overline{C}_j^{a,\overline{c}_{j-1},\overline{d}_{j-1}}, A^a\equiv a, A )$
\item for any $j=0,\ldots,K$, the natural value of the competing event indicator $D_{j+1}^{a,\overline{c}_{j+1},\overline{d}_{j}}$ and future counterfactual outcomes $\underline{Y}^{a,\overline{d}_{j+1},\overline{c}_{j+1}}_{j+1}$ conditional on $(\overline{L}^{a,\overline{c}_{j}=\overline{d}_{j}}_{j},\overline{Y}^{a,\overline{c}_{j},\overline{d}_{j}}_{j},\overline{D}_j^{a,\overline{c}_j,\overline{d}_j}\equiv d_j,\overline{D}_j^{a,\overline{c}_j,\overline{d}_{j-1}}, \overline{C}_{j+1}^{a,\overline{c}_{j+1},\overline{d}_{j}}\equiv c_{j+1}, \overline{C}_{j+1}^{a,\overline{c}_{j},\overline{d}_{j}}, A^a\equiv a, A )$
\end{enumerate}
By contrast, Figure \ref{swig2} alternatively shows the the SWIG template $\mathcal{G}(a,\overline{c},\overline{d})$ derived from the causal DAG in Figure \ref{dag3} under which the event of interest and competing event share an unmeasured common cause $U$.  Now, for example, condition 3. fails by the unblocked backdoor path $D^{a}_k\leftarrow U\rightarrow Y^{a,\overline{d}_k}_{k}$ that remains even after conditioning on $A^{a}\equiv a, A,L_0$.

\begin{figure}[tbp]
\centering
\includegraphics[scale=1]{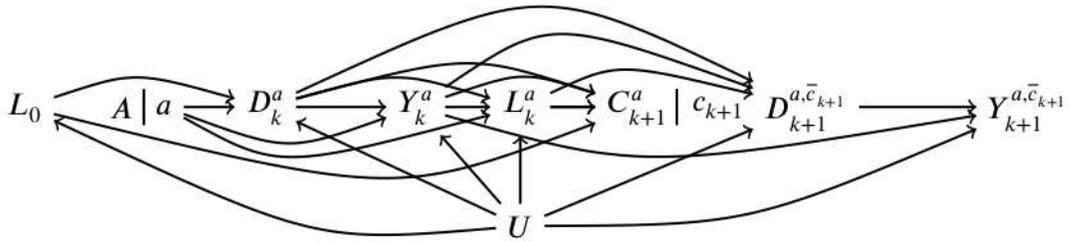}
\caption{A SWIG template $\mathcal{G}(a,\overline{c})$ derived from the causal DAG in Figure \ref{dag3} which implies exchangeability \eqref{eqn:C2} holds.  }
\label{swig3}
\end{figure}

Now consider an intervention that sets $A$ to $a$ and somehow eliminates only loss to follow-up but not competing events.  Figure \ref{swig3} is a SWIG template $\mathcal{G}(a,\overline{c})$ that is a transformation of the causal DAG in Figure \ref{dag3} for an intervention where $\overline{C}_{k+1}$ are set to fixed values.  Again, we are interested specifically in $\overline{c}=0$ (using the general template only to reduce clutter on the graph).

In this case, define $C_j^{a,\overline{c}_j}$ as the value of the censoring indicator at $j$ under an intervention that sets $\overline{C}_{K+1}=\overline{c}_{K+1}$ such that, by definition, $C_j^{a,\overline{c}_j}\equiv c_j$ for all individuals.  Then, given the consistency assumption \eqref{eqn:consistency2} and for $\overline{c}=0$, exchangeability \eqref{eqn:C2} is implied under the data generating assumptions of the causal DAG in Figure \ref{dag3} by the absence of any unblocked backdoor paths on the SWIG template $\mathcal{G}(a,\overline{c})$ (Figure \ref{swig3}) between:
 \begin{enumerate}
\item the natural value of treatment $A$ and future counterfactual outcomes $\overline{Y}^{a,\overline{c}}_{K+1}$ conditional on $L_0$
\item for any $j=0,\ldots,K$, the natural value of the loss to follow-up indicator $C_{j+1}^{a,\overline{c}_{j}}$ and future counterfactual outcomes $\underline{Y}^{a,\overline{c}_{j+1}}_{j+1}$ conditional on $(\overline{L}^{a,\overline{c}_{j}}_{j},\overline{Y}^{a,\overline{c}_{j}}_{j},\overline{D}_j^{a,\overline{c}_j}, \overline{C}_j^{a,\overline{c}_j}\equiv c_j, \overline{C}_j^{a,\overline{c}_{j-1}}, A^a\equiv a, A )$
 \end{enumerate}

\section{Identification Proofs}\label{appb}
For simplicity and without loss of generality, in this section take $L_0$ to be a constant (i.e. we consider a population restricted to one level of $L_0$) and thus all estimands and assumptions can be implicitly interpreted as conditional on $L_0$).

\begin{customthm}{1}\label{th1}
Suppose the following identifying assumptions hold:

\begin{enumerate}
\item Exchangeability:
\begin{equation*}
\overline{Y}_{K+1}^{a,\overline{v}=0}\coprod A,
\end{equation*}
\begin{equation}
\underline{Y}_{k+1}^{a,\overline{v}=0}\coprod V_{k+1}|%
\overline{X}_k=\overline{x}_k,\overline{Y}_k=\overline{V}_k=0,A=a  \label{eqn:C}
\end{equation}

\item Positivity:
\begin{align}
&f_{A,\overline{X}_{k},V_k,Y_k}(a,\overline{x}_{k},0,0)\neq0\implies\nonumber\\
&\Pr[V_{k+1}=0|\overline{X}_k=\overline{x}_{k},V_k=Y_k=0,A=a]>0\mbox{ w.p.1}. \label{eqn:positivity}
\end{align}

\item Consistency:
\begin{align}
&\mbox{If } A=a \mbox{ and }\overline{V}_{k+1}=0\mbox{,} \nonumber\\
&\mbox{ then } \overline{X}_{k+1}=\overline{X}_{k+1}^{a,\overline{v}=0} \mbox{ and } \overline{Y}_{k+1}=\overline{Y}_{k+1}^{a,\overline{v}=0} \label{eqn:consistency}
\end{align}
\end{enumerate}
with $V_{k+1}$ a vector of censoring indicators including all sources of censoring by $k+1$ (with $V_0\equiv 0$) and $X_k$ a vector of measured time-varying covariates at $k$.  Then $\Pr [Y_{K+1}^{a,\overline{v}=\overline{0}}=1]$, the counterfactual risk under $a$ and elimination of censoring, is equivalent to 
\begin{align}
\sum_{\overline{x}_{K}}\sum_{k=0}^{K}&\Pr [Y_{k+1}=1|\overline{X}_{k}=%
\overline{x}_{k},\overline{V}_{k+1}=\overline{Y}%
_{k}=0,A=a]\times  \notag \\
&\prod_{j=0}^{k}\{ \Pr [Y_{j}=0|\overline{X}_{j-1}=\overline{x}_{j-1},%
\overline{V}_{j}=\overline{Y}_{j-1}=0,A=a]\times  \notag \\
& f(x_{j}|\overline{x}_{j-1},\overline{V}_{j}=\overline{Y}%
_{j}=0,a)\}   \label{eqn:gform}
\end{align}
\end{customthm}
Proof: For $k=0,\ldots,K$ and $Y_{K+1}^{a,\overline{v}=0}\equiv Y^{a,\overline{v}=0}$, if $f(\overline{X}_k,V_k=Y_k=0,a)\neq0$ then define $b^{a,\overline{v}=0}(K,\overline{X}_k)=\Pr[Y^{a,\overline{v}=0}=0|\overline{X}_k,\overline{V}_k=\overline{Y}_k=0,a]$.  For $V_0\equiv Y_0\equiv0$ and $X_0$ a constant, we have $b^{a,\overline{v}=0}(K,X_0)=\Pr[Y^{a,\overline{v}=0}=0|A=a]$.  By \eqref{eqn:C}, we further have $b^{a,\overline{v}=0}(K,X_0)=\Pr[Y^{a,\overline{v}=0}=0]$.

The event $Y^{a,\overline{v}=0}=0$ implies $Y_k^{a,\overline{v}=0}=0$, $k=0,\ldots,K$.  Therefore, $b^{a,\overline{v}=0}(K,\overline{X}_k)=\Pr[Y_{k+1}^{a,\overline{v}=0},\ldots,Y_{K+1}^{a,\overline{v}=0}=0|\overline{X}_k,\overline{V}_{k}=\overline{Y}_k=0,a]$ for any $k$.  

By \eqref{eqn:C} and \eqref{eqn:positivity}
\begin{equation*}
b^{a,\overline{v}=0}(K,\overline{X}_k)=\Pr[Y_{k+1}^{a,\overline{v}=0},\ldots,Y_{K+1}^{a,\overline{v}=0}=0|\overline{X}_k,\overline{V}_{k+1}=\overline{Y}_k=0,a]
\end{equation*}

Further by \eqref{eqn:consistency} and laws of probability 
\begin{align*}
b^{a,\overline{v}=0}(K,\overline{X}_k)&=\Pr[Y_{k+2}^{a,\overline{v}=0},\ldots,Y_{K+1}^{a,\overline{v}=0}=0|\overline{X}_k,\overline{V}_{k+1}=\overline{Y}_{k+1}=0,a]\\
&\times\Pr[Y_{k+1}=0|\overline{X}_k,\overline{V}_{k+1}=\overline{Y}_{k}=0,a]\\
&=\sum_{x_{k+1}}\Pr[Y_{k+2}^{a,\overline{v}=0},\ldots,Y_{K+1}^{a,\overline{v}=0}=0|X_{k+1}=x_{k+1},\overline{X}_k,\overline{V}_{k+1}=\overline{Y}_{k+1}=0,a]\\
&\times\Pr[Y_{k+1}=0|\overline{X}_k,\overline{V}_{k+1}=\overline{Y}_{k}=0,a]f(x_{k+1}|\overline{X}_k,\overline{V}_{k+1}=\overline{Y}_{k+1}=0,a)
\end{align*}

By another invocation of \eqref{eqn:C} and \eqref{eqn:positivity}
\begin{align*}
b^{a,\overline{v}=0}(K,\overline{X}_k)&=\sum_{x_{k+1}}\Pr[Y_{k+2}^{a,\overline{v}=0},\ldots,Y_{K+1}^{a,\overline{v}=0}=0|X_{k+1}=x_{k+1},\overline{X}_k,\overline{V}_{k+2}=\overline{Y}_{k+1}=0,a]\\
&\times\Pr[Y_{k+1}=0|\overline{X}_k,\overline{V}_{k+1}=\overline{Y}_{k}=0,a]f(x_{k+1}|\overline{X}_k,\overline{V}_{k+1}=\overline{Y}_{k+1}=0,a)
\end{align*}

Arguing iteratively, for $k=0,\ldots,K$ we have
\begin{align*}
b^{a,\overline{v}=0}(K,\overline{X}_k)&=\sum_{x_{k+1}}\ldots \sum_{x_{K}}\Pr[Y_{K+1}^{a,\overline{v}=0}=0|X_{k+1}=x_{k+1},\ldots, X_K=x_K,\overline{X}_k,\overline{V}_{K+1}=\overline{Y}_{K}=0,a]\\
&\times \Pr[Y_{k+1}=0|\overline{X}_k,\overline{V}_{k+1}=\overline{Y}_{k}=0,a],\ldots, \\
&\times \Pr[Y_{K}=0|X_{k+1}=x_{k+1},\ldots,X_{K-1}=x_{k-1},\overline{X}_k,\overline{V}_{K}=\overline{Y}_{K-1}=0,a]\\
&\times f(x_{k+1}|\overline{X}_k,\overline{V}_{k+1}=\overline{Y}_{k+1}=0,a),\ldots,f(x_{K}|x_{k+1},\ldots,x_{K-1},\overline{X}_k,\overline{V}_{K}=\overline{Y}_{K}=0,a)
\end{align*}

Setting $k=0$, noting that $Y_0\equiv 0$ by definition, and invoking consistency \eqref{eqn:consistency} once more we have
\begin{align*}
b^{a,\overline{v}=0}(K,X_0)=&\sum_{\overline{x}_K}\Pr[Y_{K+1}=0|\overline{X}_K=\overline{x}_K,\overline{V}_{K+1}=\overline{Y}_{K}=0,a]\\
&\prod_{j=0}^{K}\Pr[Y_{j}=0|\overline{X}_{j-1}=\overline{x}_{j-1},\overline{V}_{j}=\overline{Y}_{j-1}=0,a]f(x_{j}|\overline{x}_{j-1},\overline{V}_{j}=\overline{Y}_{j}=0,a)
\end{align*}
The result follows by noting that the complement of $b^{a,\overline{v}=0}(K,X_0)$ is equal to \eqref{eqn:gform}.

\begin{customcorollary}{1}\label{corr1}
Given \eqref{eqn:C1} , \eqref{eqn:positivity1} and \eqref{eqn:consistency1}, the counterfactual risk under elimination of competing events \eqref{eqn:risk1} equals the g-formula \eqref{eqn:gform1}.
\end{customcorollary}
Proof:  The result follows from Theorem \ref{th1} by choosing $\overline{V}_{K+1}=(\overline{C}_{K+1},\overline{D}_{K+1})$ and $\overline{X}_K=\overline{L}_K$.

\begin{customcorollary}{2}\label{corr2}
Given \eqref{eqn:C2} , \eqref{eqn:positivity2} and \eqref{eqn:consistency2}, the counterfactual risk without elimination of competing events \eqref{eqn:risk2} equals the g-formula \eqref{eqn:gform2}.
\end{customcorollary}
Proof:  Assumptions \eqref{eqn:C2} , \eqref{eqn:positivity2} and \eqref{eqn:consistency2} are special cases of \eqref{eqn:C} , \eqref{eqn:positivity} and \eqref{eqn:consistency} choosing $\overline{V}_{K+1}=\overline{C}_{K+1}$ and $\overline{X}_K=(\overline{L}_K,\overline{D}_K)$.  By Theorem \ref{th1}, given these assumptions we have that $\Pr[Y_{K+1}^{a,\overline{c}=0}=0]$  equals
\begin{align*}
\sum_{\overline{l}_K}\sum_{\overline{d}_K}&\Pr[Y_{K+1}=0|\overline{L}_K=\overline{l}_K,\overline{D}_K=\overline{d}_K,\overline{C}_{K+1}=\overline{Y}_{K}=0,a]\\
&\prod_{j=0}^{K}\Pr[Y_{j}=0|\overline{L}_{j-1}=\overline{l}_{j-1},\overline{D}_{j-1}=\overline{d}_{j-1},\overline{C}_{j}=\overline{Y}_{j-1}=0,a]\\
&\times f(l_{j},d_j|\overline{l}_{j-1},\overline{d}_{j-1},\overline{C}_{j}=\overline{Y}_{j}=0,a)
\end{align*}
which, by laws of probability, can be equivalently written as:
\begin{align*}
\sum_{\overline{l}_K}\sum_{\overline{d}_K}&\Pr[Y_{K+1}=0|\overline{L}_K=\overline{l}_K,\overline{D}_K=\overline{d}_K,\overline{C}_{K+1}=\overline{Y}_{K}=0,a]\\
&\prod_{j=0}^{K}\Pr[Y_{j}=0|\overline{L}_{j-1}=\overline{l}_{j-1},\overline{D}_{j}=\overline{d}_{j},\overline{C}_{j}=\overline{Y}_{j-1}=0,a]\\
&\times \Pr[D_{j}=d_j|\overline{l}_{j-1},\overline{d}_{j-1}\overline{C}_{j}=\overline{Y}_{j-1}=0,a]
f(l_{j}|\overline{l}_{j-1},\overline{d}_{j},\overline{C}_{j}=\overline{Y}_{j}=0,a)\\
&=\\
\sum_{\overline{l}_K}\sum_{\overline{d}_{K+1}}&\Pr[Y_{K+1}=0,D_{K+1}=d_{K+1}|\overline{L}_K=\overline{l}_K,\overline{D}_K=\overline{d}_K,\overline{C}_{K+1}=\overline{Y}_{K}=0,a]\\
&\prod_{j=0}^{K}\Pr[Y_{j}=0|\overline{L}_{j-1}=\overline{l}_{j-1},\overline{D}_{j}=\overline{d}_{j},\overline{C}_{j}=\overline{Y}_{j-1}=0,a]\\
&\times \Pr[D_{j}=d_j|\overline{l}_{j-1},\overline{d}_{j-1}\overline{C}_{j}=\overline{Y}_{j-1}=0,a]
f(l_{j}|\overline{l}_{j-1},\overline{d}_{j},\overline{C}_{j}=\overline{Y}_{j}=0,a)\\
&=\\
\sum_{\overline{l}_K}\sum_{\overline{d}_{K+1}}&\Pr[Y_{K+1}=0|\overline{L}_K=\overline{l}_K,\overline{D}_{K+1}=\overline{d}_{K+1},\overline{C}_{K+1}=\overline{Y}_{K}=0,a]\\
&\prod_{j=0}^{K}\Pr[Y_{j}=0|\overline{L}_{j-1}=\overline{l}_{j-1},\overline{D}_{j}=\overline{d}_{j},\overline{C}_{j}=\overline{Y}_{j-1}=0,a]\\
&\times \Pr[D_{j+1}=d_{j+1}|\overline{l}_{j},\overline{d}_{j}\overline{C}_{j+1}=\overline{Y}_{j}=0,a]
f(l_{j}|\overline{l}_{j-1},\overline{d}_{j},\overline{C}_{j}=\overline{Y}_{j}=0,a)\\
\end{align*}
The last expression is the complement of \eqref{eqn:gform2}.

\begin{customlemma}{1}\label{lemma1}
Defining  
\begin{equation}
h^{\overline{v}=0}_{k}(a)=\frac{\mbox{E}\lbrack Y_{k+1}(1-Y_{k})W^{V}_{k}|A=a]}{\mbox{E}\lbrack
(1-Y_{k})W^{V}_{k}]A=a]}  \label{eqn:h1v}
\end{equation}
and 
\begin{equation}
W^{V}_k=\prod_{j=0}^{k}\frac{I(V_{j+1}=0)}{\Pr[V_{j+1}=0|%
\overline{X}_j,\overline{V}_{j}=\overline{Y}%
_{j}=0,A=a]}  \label{eqn:wv}
\end{equation}%
we have, for $k=0,\ldots,K$,
\begin{align*}
&h^{\overline{v}=0}_{k}(a)\prod_{j=0}^{k-1}\left[1-h^{\overline{v}=0}_{j}(a)\right]=\\
&\mbox{E}[Y_{k+1}(1-Y_k)W_k^V|A=a]\prod_{j=1}^k\frac{\mbox{E}[(1-Y_{j-1})W_{j-1}^V|A=a]-\mbox{E}[Y_j(1-Y_{j-1})W_{j-1}^V|A=a]}{\mbox{E}[(1-Y_{j})W_{j}^V|A=a]}
\end{align*}
\end{customlemma}
Proof: For some $0\leq t<k-1$, by \eqref{eqn:h1v} we have
\begin{align*}
&h^{\overline{v}=0}_{k}(a)\prod_{j=t}^{k-1}\left[1-h^{\overline{v}=0}_{j}(a)\right]=\\
&\frac{\mbox{E}[Y_{k+1}(1-Y_k)W_k^V|A=a]}{\mbox{E}[(1-Y_k)W_k^V|A=a]}\times \left\{1-\frac{\mbox{E}[Y_{k}(1-Y_{k-1})W_{k-1}^V|A=a]}{\mbox{E}[(1-Y_{k-1})W_{k-1}^V|A=a]} \right\}\times \ldots \times \\
&\left\{1-\frac{\mbox{E}[Y_{t+1}(1-Y_t)W_t^V|A=a]}{\mbox{E}[(1-Y_t)W_t^V|A=a]} \right\}=\\
&\frac{\mbox{E}[Y_{k+1}(1-Y_k)W_k^V|A=a]}{\mbox{E}[(1-Y_k)W_k^V|A=a]}\times \left\{\frac{\mbox{E}[(1-Y_{k-1})W_{k-1}^V|A=a]-\mbox{E}[Y_{k}(1-Y_{k-1})W_{k-1}^V|A=a]}{\mbox{E}[(1-Y_{k-1})W_{k-1}^V|A=a]} \right\}\times \ldots \times \\
&\left\{\frac{\mbox{E}[(1-Y_t)W_t^V|A=a]-\mbox{E}[Y_{t+1}(1-Y_t)W_t^V|A=a]}{\mbox{E}[(1-Y_t)W_t^V|A=a]} \right\}=\\
&\mbox{E}[Y_{k+1}(1-Y_k)W_k^V|A=a] \times \left\{\frac{\mbox{E}[(1-Y_{k-1})W_{k-1}^V|A=a]-\mbox{E}[Y_{k}(1-Y_{k-1})W_{k-1}^V|A=a]}{\mbox{E}[(1-Y_k)W_k^V|A=a]} \right\}\times\\
&\left\{\frac{\mbox{E}[(1-Y_{k-2})W_{k-2}^V|A=a]-\mbox{E}[Y_{k-1}(1-Y_{k-2})W_{k-2}^V|A=a]}{\mbox{E}[(1-Y_{k-1})W_{k-1}^V|A=a]} \right\}\times \ldots \times \\
&\left\{\frac{\mbox{E}[(1-Y_t)W_t^V|A=a]-\mbox{E}[Y_{t+1}(1-Y_t)W_t^V|A=a]}{\mbox{E}[(1-Y_{t+1})W_{t+1}^V|A=a]} \right\}\times \frac{1}{\mbox{E}[(1-Y_{t})W_{t}^V|A=a]}=\\
&\mbox{E}[Y_{k+1}(1-Y_k)W_k^V|A=a]\left\{\prod_{j=t+1}^k\frac{\mbox{E}[(1-Y_{j-1})W_{j-1}^V|A=a]-\mbox{E}[Y_j(1-Y_{j-1})W_{j-1}^V|A=a]}{\mbox{E}[(1-Y_{j})W_{j}^V|A=a]}\right\}\times\\
& \frac{1}{\mbox{E}[(1-Y_{t})W_{t}^V|A=a]}
\end{align*}
Our result follows by setting $t=0$ and noting that $\mbox{E}[(1-Y_{0})W_{0}^V|A=a]=1$.

\begin{customlemma}{2}\label{lemma2}
For $k=0,\ldots,K$ and $W_k^V$ as in \eqref{eqn:wv}
\begin{equation*}
\mbox{E}[(1-Y_k)W_k^V|A=a]=\mbox{E}[(1-Y_{k-1})W_{k-1}^V|A=a]-\mbox{E}[Y_k(1-Y_{k-1})W_{k-1}^V|A=a]
\end{equation*}
\end{customlemma}
Proof: By laws of expectation
\begin{align*}
&\mbox{E}[(1-Y_k)W_k^V|A=a]=\\
&\mbox{E}\left [(1-Y_k)W_{k-1}^V\mbox{E}\left [\frac{I(V_{k+1}=0)}{\Pr[V_{k+1}=0|%
\overline{X}_k,\overline{V}_{k}=\overline{Y}%
_{k}=0,A=a]}| \overline{X}_k,\overline{V}_k,\overline{Y}_k\right ]|A=a\right ]=\\
&\mbox{E}[(1-Y_k)W_{k-1}^V|A=a]
\end{align*}
As the event $Y_k=0$ implies the joint event $(Y_k=0,Y_{k-1}=0)$ we have
\begin{align*}
&\mbox{E}[(1-Y_k)W_{k-1}^V|A=a]=\mbox{E}[(1-Y_k)(1-Y_{k-1})W_{k-1}^V|A=a]=\\
&\mbox{E}[(1-Y_{k-1})W_{k-1}^V-Y_k(1-Y_{k-1})W_{k-1}^V|A=a]=\\
&\mbox{E}[(1-Y_{k-1})W_{k-1}^V|A=a]-\mbox{E}[Y_k(1-Y_{k-1})W_{k-1}^V|A=a]
\end{align*}

\begin{customlemma}{3}\label{lemma3}
For each $k=0,\ldots,K$
\begin{align}
&\mbox{E}[Y_{k+1}(1-Y_{k})W_{k}^V|A=a]=\nonumber\\
&\sum_{\overline{x}_k}\Pr[Y_{k+1}=1|\overline{X}_k=\overline{x}_k,\overline{V}_{k+1}=\overline{Y}_{k}=0,a]\times\nonumber\\
&\prod_{j=0}^{k}\Pr[Y_{j}=0|\overline{X}_{j-1}=\overline{x}_{j-1},\overline{V}_{j}=\overline{Y}_{j-1}=0,a]f(x_{j}|\overline{x}_{j-1},\overline{V}_{j}=\overline{Y}_{j}=0,a)\label{eqn:expfail}
\end{align}
and 
\begin{align}
&\mbox{E}[(1-Y_{k})W_{k}^V|A=a]=\nonumber\\
&\sum_{\overline{x}_{k-1}}\Pr[Y_{k}=0|\overline{X}_{k-1}=\overline{x}_{k-1},\overline{V}_{k}=\overline{Y}_{k-1}=0,a]\times\nonumber\\
&\prod_{j=0}^{k-1}\Pr[Y_{j}=0|\overline{X}_{j-1}=\overline{x}_{j-1},\overline{V}_{j}=\overline{Y}_{j-1}=0,a]f(x_{j}|\overline{x}_{j-1},\overline{V}_{j}=\overline{Y}_{j}=0,a)\label{eqn:expsurv}
\end{align}
\end{customlemma}
Proof: By definition we have
\begin{align*}
&\mbox{E}[Y_{k+1}(1-Y_{k})W_{k}^V|A=a]=\\
&\sum_{\overline{y}_{k+1}}\sum_{\overline{x}_k}\sum_{\overline{v}_{k+1}}y_{k+1}(1-y_k)\prod_{j=0}^k\frac{(1-v_{j+1})}{\Pr[V_{j+1}=0|%
\overline{x}_j,\overline{V}_{j}=\overline{Y}%
_{j}=0,A=a]}f(y_{j+1}|\overline{y}_j,\overline{x}_j,\overline{v}_{j+1},a)\\
&f(v_{j+1}|\overline{y}_j,\overline{x}_j,\overline{v}_j,a)f(x_j|\overline{y}_j,\overline{v}_j,\overline{x}_{j-1},a)
\end{align*}
The result \eqref{eqn:expfail} follows by noting that any component of the above sum will be zero when $y_{k+1}=0$, $y_j=1$ or $v_{j+1}=1$, $j=0,\ldots,k$.  Analogous arguments prove \eqref{eqn:expsurv}.

\begin{customthm}{2}\label{th2}
Expression \eqref{eqn:gform} equals
\begin{equation}
\sum_{k=0}^{K}h^{\overline{v}=0}_{k}(a)\prod_{j=0}^{k-1}\left[1-h^{\overline{v}=0}_{j}(a)\right]  \label{eqn:ipw}
\end{equation}
\end{customthm}
Proof: By Lemma \ref{lemma1}, we can rewrite \eqref{eqn:ipw} as
\begin{equation*}
\sum_{k=0}^{K}\mbox{E}[Y_{k+1}(1-Y_k)W_k^V|A=a]\prod_{j=1}^k\frac{\mbox{E}[(1-Y_{j-1})W_{j-1}^V|A=a]-\mbox{E}[Y_j(1-Y_{j-1})W_{j-1}^V|A=a]}{\mbox{E}[(1-Y_{j})W_{j}^V|A=a]}
\end{equation*}
which, by Lemma \ref{lemma2}, reduces to
\begin{equation*}
\sum_{k=0}^{K}\mbox{E}[Y_{k+1}(1-Y_k)W_k^V|A=a]
\end{equation*}
The result then follows from Lemma \ref{lemma3}.

\begin{customcorollary}{3}\label{corr3}
The g-formula \eqref{eqn:gform1} equals expression \eqref{eqn:ipw1}.
\end{customcorollary}
The result follows from Theorem \ref{th2} by choosing $\overline{V}_{K+1}=(\overline{C}_{K+1},\overline{D}_{K+1})$ and $\overline{X}_K=\overline{L}_K$.

\begin{customcorollary}{4}\label{corr4}
The g-formula \eqref{eqn:gform2} equals expression \eqref{eqn:ipw2first}.
\end{customcorollary}
The result follows from Theorem \ref{th2} by choosing $\overline{V}_{K+1}=\overline{C}_{K+1}$ and $\overline{X}_K=(\overline{L}_K,\overline{D}_K)$.

\begin{customlemma}{4}\label{lemma4}
Given the definitions \eqref{eqn:cs1} and \eqref{eqn:cs2} in the main text:
\begin{align*}
&h^{1}
_{k}(a)\{1-h^{2}_{k}(a)\}\prod_{j=t}^{k-1}[\{1-h^{1}_{j}(a)\}\{1-h^{2}_{j}(a)\}]=\\
&\mbox{E}[Y_{k+1}(1-D_{k+1})(1-Y_k)W_k^{C}|A=a]\times\\
&\prod_{j=0}^k\frac{\mbox{E}[(1-Y_{j})(1-D_j)W_{j}^C|A=a]-\mbox{E}[D_{j+1}(1-Y_{j})(1-D_j)W_{j}^C|A=a]}{\mbox{E}[(1-D_{j+1})(1-Y_j)W_{j}^C|A=a]}\times\\
&\prod_{j=1}^k\frac{\mbox{E}[(1-D_{j})(1-Y_{j-1})W_{j-1}^C|A=a]-\mbox{E}[Y_{j}(1-D_{j})(1-Y_{j-1})W_{j-1}^C|A=a]}{\mbox{E}[(1-Y_{j})(1-D_j)W_{j}^C|A=a]}
\end{align*}
\end{customlemma}
Proof: For some $0\leq t<k-1$ by \eqref{eqn:cs1} and \eqref{eqn:cs2} in the main text we have
\begin{align*}
&h^{1}
_{k}(a)\{1-h^{2}_{k}(a)\}\prod_{j=t}^{k-1}[\{1-h^{1}_{j}(a)\}\{1-h^{2}_{j}(a)\}]=\\
&\frac{\mbox{E}[Y_{k+1}(1-D_{k+1})(1-Y_k)W_k^C|A=a]}{\mbox{E}[(1-D_{k+1})(1-Y_k)W_k^C|A=a]} \left\{1-\frac{\mbox{E}[D_{k+1}(1-Y_{k})(1-D_k)W_{k}^C|A=a]}{\mbox{E}[(1-Y_{k})(1-D_k)W_{k}^C|A=a]} \right\}\times\\
& \left\{1-\frac{\mbox{E}[Y_{k}(1-D_{k})(1-Y_{k-1})W_{k-1}^C|A=a]}{\mbox{E}[(1-D_{k})(1-Y_{k-1})W_{k-1}^C|A=a]} \right\} \left\{1-\frac{\mbox{E}[D_{k}(1-Y_{k-1})(1-D_{k-1})W_{k-1}^C|A=a]}{\mbox{E}[(1-Y_{k-1})(1-D_{k-1})W_{k-1}^C|A=a]} \right\}\times \ldots \times\\
& \left\{1-\frac{\mbox{E}[Y_{t+1}(1-D_{t+1})(1-Y_{t})W_{t}^C|A=a]}{\mbox{E}[(1-D_{t+1})(1-Y_{t})W_{t}^C|A=a]} \right\} \left\{1-\frac{\mbox{E}[D_{t+1}(1-Y_{t})(1-D_{t})W_{t}^C|A=a]}{\mbox{E}[(1-Y_{t})(1-D_{t})W_{t}^C|A=a]} \right\}=\\
\\
&\frac{\mbox{E}[Y_{k+1}(1-D_{k+1})(1-Y_k)W_k^C|A=a]}{\mbox{E}[(1-D_{k+1})(1-Y_k)W_k^C|A=a]}\times\\
& \frac{\mbox{E}[(1-Y_{k})(1-D_k)W_{k}^C|A=a]-\mbox{E}[D_{k+1}(1-Y_{k})(1-D_k)W_{k}^C|A=a]}{\mbox{E}[(1-Y_{k})(1-D_k)W_{k}^C|A=a]} \times\\
&\frac{\mbox{E}[(1-D_{k})(1-Y_{k-1})W_{k-1}^C|A=a]-\mbox{E}[Y_{k}(1-D_{k})(1-Y_{k-1})W_{k-1}^C|A=a]}{\mbox{E}[(1-D_{k})(1-Y_{k-1})W_{k-1}^C|A=a]} \times\\
&\frac{\mbox{E}[(1-Y_{k-1})(1-D_{k-1})W_{k-1}^C|A=a]-\mbox{E}[D_{k}(1-Y_{k-1})(1-D_{k-1})W_{k-1}^C|A=a]}{\mbox{E}[(1-Y_{k-1})(1-D_{k-1})W_{k-1}^C|A=a]} \times \ldots \times\\
&\frac{\mbox{E}[(1-D_{t+1})(1-Y_{t})W_{t}^C|A=a]-\mbox{E}[Y_{t+1}(1-D_{t+1})(1-Y_{t})W_{t}^C|A=a]}{\mbox{E}[(1-D_{t+1})(1-Y_{t})W_{t}^C|A=a]} \times\\
&\frac{\mbox{E}[(1-Y_{t})(1-D_{t})W_{t}^C|A=a]-\mbox{E}[D_{t+1}(1-Y_{t})(1-D_{t})W_{t}^C|A=a]}{\mbox{E}[(1-Y_{t})(1-D_{t})W_{t}^C|A=a]} =\\
\\
&\mbox{E}[Y_{k+1}(1-D_{k+1})(1-Y_k)W_k^C|A=a]\times\\
& \frac{\mbox{E}[(1-Y_{k})(1-D_k)W_{k}^C|A=a]-\mbox{E}[D_{k+1}(1-Y_{k})(1-D_k)W_{k}^C|A=a]}{\mbox{E}[(1-D_{k+1})(1-Y_k)W_k^C|A=a]} \times\\
&\frac{\mbox{E}[(1-D_{k})(1-Y_{k-1})W_{k-1}^C|A=a]-\mbox{E}[Y_{k}(1-D_{k})(1-Y_{k-1})W_{k-1}^C|A=a]}{\mbox{E}[(1-Y_{k})(1-D_k)W_{k}^C|A=a]} \times\\
&\frac{\mbox{E}[(1-Y_{k-1})(1-D_{k-1})W_{k-1}^C|A=a]-\mbox{E}[D_{k}(1-Y_{k-1})(1-D_{k-1})W_{k-1}^C|A=a]}{\mbox{E}[(1-D_{k})(1-Y_{k-1})W_{k-1}^C|A=a]} \times\\
&\frac{\mbox{E}[(1-D_{k-1})(1-Y_{k-2})W_{k-2}^C|A=a]-\mbox{E}[Y_{k-1}(1-D_{k-1})(1-Y_{k-2})W_{k-2}^C|A=a]}{\mbox{E}[(1-Y_{k-1})(1-D_{k-1})W_{k-1}^C|A=a]} \times  \ldots \times\\
&\frac{\mbox{E}[(1-D_{t+1})(1-Y_{t})W_{t}^C|A=a]-\mbox{E}[Y_{t+1}(1-D_{t+1})(1-Y_{t})W_{t}^C|A=a]}{\mbox{E}[(1-Y_{t+1})(1-D_{t+1})W_{t+1}^C|A=a]} \times\\
&\frac{\mbox{E}[(1-Y_{t})(1-D_{t})W_{t}^C|A=a]-\mbox{E}[D_{t+1}(1-Y_{t})(1-D_{t})W_{t}^C|A=a]}{\mbox{E}[(1-D_{t+1})(1-Y_{t})W_{t}^C|A=a]}\times\\
& \frac{1}{\mbox{E}[(1-Y_{t})(1-D_{t})W_{t}^C|A=a]} 
\end{align*}
Thus
\begin{align*}
&h^{1}
_{k}(a)\{1-h^{2}_{k}(a)\}\prod_{j=t}^{k-1}[\{1-h^{1}_{j}(a)\}\{1-h^{2}_{j}(a)\}]=\\
&\mbox{E}[Y_{k+1}(1-D_{k+1})(1-Y_k)W_k^{C}|A=a]\times\\
&\prod_{j=t}^k\frac{\mbox{E}[(1-Y_{j})(1-D_j)W_{j}^C|A=a]-\mbox{E}[D_{j+1}(1-Y_{j})(1-D_j)W_{j}^C|A=a]}{\mbox{E}[(1-D_{j+1})(1-Y_j)W_{j}^C|A=a]}\times\\
&\prod_{j=t+1}^k\frac{\mbox{E}[(1-D_{j})(1-Y_{j-1})W_{j-1}^C|A=a]-\mbox{E}[Y_{j}(1-D_{j})(1-Y_{j-1})W_{j-1}^C|A=a]}{\mbox{E}[(1-Y_{j})(1-D_j)W_{j}^C|A=a]}\times\\
& \frac{1}{\mbox{E}[(1-Y_{t})(1-D_{t})W_{t}^C|A=a]} 
\end{align*}
The result follows by setting $t=0$ and noting $\mbox{E}[(1-Y_{0})(1-D_{0})W_{0}^C|A=a]=1$

\begin{customlemma}{5}\label{lemma5}
For any follow-up interval $j$
\begin{equation*}
\mbox{E}[(1-D_{j+1})(1-Y_{j})W_{j}^C|A=a]=\mbox{E}[(1-Y_{j})(1-D_{j})W_{j}^C|A=a]-\mbox{E}[D_{j+1}(1-Y_{j})(1-D_{j})W_{j}^C|A=a]
\end{equation*}
and
\begin{equation*}
\mbox{E}[(1-Y_{j})(1-D_{j})W_{j}^C|A=a]=\mbox{E}[(1-D_{j})(1-Y_{j-1})W_{j-1}^C|A=a]-\mbox{E}[Y_{j}(1-D_{j})(1-Y_{j-1})W_{j-1}^C|A=a]
\end{equation*}
\end{customlemma}
Proof: The event $(D_{j+1}=Y_j=0)$ implies the event $(D_{j+1}=Y_j=D_j=0)$.  Therefore
\begin{align*}
&\mbox{E}[(1-D_{j+1})(1-Y_{j})W_{j}^C|A=a]=\mbox{E}[(1-D_{j+1})(1-Y_{j})(1-D_j)W_{j}^C|A=a]=\\
&\mbox{E}[(1-Y_{j})(1-D_j)W_{j}^C-D_{j+1}(1-Y_{j})(1-D_j)W_{j}^C|A=a]=\\
&\mbox{E}[(1-Y_{j})(1-D_j)W_{j}^C|A=a]-\mbox{E}[D_{j+1}(1-Y_{j})(1-D_j)W_{j}^C|A=a]
\end{align*}

Also
\begin{align*}
&\mbox{E}[(1-Y_{j})(1-D_{j})W_{j}^C|A=a]=\\
&\mbox{E}\left[(1-Y_{j})(1-D_{j})W_{j-1}^C\frac{I(C_{j+1}=0)}{\Pr[C_{j+1}=0|\overline{L}_j,\overline{C}_j=\overline{D}_j=\overline{Y}_j=0,A=a]}|A=a\right]=\\
&\mbox{E}\left[\mbox{E}\left[(1-Y_{j})(1-D_{j})W_{j-1}^C\frac{I(C_{j+1}=0)}{\Pr[C_{j+1}=0|\overline{L}_j,\overline{C}_j=\overline{D}_j=\overline{Y}_j=0,A=a]}|\overline{L}_j,\overline{C}_j,\overline{D}_j,\overline{Y}_j\right]|A=a\right]=\\
&\mbox{E}\left[(1-Y_{j})(1-D_{j})W_{j-1}^C\mbox{E}\left[\frac{I(C_{j+1}=0)}{\Pr[C_{j+1}=0|\overline{L}_j,\overline{C}_j=\overline{D}_j=\overline{Y}_j=0,A=a]}|\overline{L}_j,\overline{C}_j,\overline{D}_j,\overline{Y}_j\right]|A=a\right]=\\
&\mbox{E}\left[(1-Y_{j})(1-D_{j})W_{j-1}^C|A=a\right]
\end{align*}
Further, the event $(Y_j=D_j=0)$ implies the event $(Y_j=D_j=Y_{j-1}=0)$ such that
\begin{align*}
&\mbox{E}[(1-Y_{j})(1-D_{j})W_{j-1}^C|A=a]=\mbox{E}[(1-Y_{j})(1-D_{j})(1-Y_{j-1})W_{j-1}^C|A=a]=\\
&\mbox{E}[(1-D_{j})(1-Y_{j-1})W_{j-1}^C-Y_j(1-D_{j})(1-Y_{j-1})W_{j-1}^C|A=a]=\\
&\mbox{E}[(1-D_{j})(1-Y_{j-1})W_{j-1}^C|A=a]-\mbox{E}[Y_j(1-D_{j})(1-Y_{j-1})W_{j-1}^C|A=a]
\end{align*}

\begin{customthm}{3}\label{th3}
The g-formula \eqref{eqn:gform2taub} is equivalent to expression \eqref{eqn:ipw2}.
\end{customthm}
Proof: By Lemma \ref{lemma4}, we can write \eqref{eqn:ipw2} as
\begin{align}
&\sum_{k=0}^{K}\mbox{E}[Y_{k+1}(1-D_{k+1})(1-Y_k)W_k^{C}|A=a]\times\nonumber\\
&\prod_{j=0}^k\frac{\mbox{E}[(1-Y_{j})(1-D_j)W_{j}^C|A=a]-\mbox{E}[D_{j+1}(1-Y_{j})(1-D_j)W_{j}^C|A=a]}{\mbox{E}[(1-D_{j+1})(1-Y_j)W_{j}^C|A=a]}\times\nonumber\\
&\prod_{j=1}^k\frac{\mbox{E}[(1-D_{j})(1-Y_{j-1})W_{j-1}^C|A=a]-\mbox{E}[Y_{j}(1-D_{j})(1-Y_{j-1})W_{j-1}^C|A=a]}{\mbox{E}[(1-Y_{j})(1-D_j)W_{j}^C|A=a]}\label{eqn:t31}
\end{align}

By Lemma \ref{lemma5}, \eqref{eqn:t31} reduces to
\begin{equation}
\sum_{k=0}^{K}\mbox{E}[Y_{k+1}(1-D_{k+1})(1-Y_k)W_k^{C}|A=a]\label{eqn:t32}
\end{equation}
For each index $k$ in the sum in \eqref{eqn:t32} we have
\begin{align*}
&\mbox{E}[Y_{k+1}(1-D_{k+1})(1-Y_k)W_k^{C}|A=a]=\\
&\sum_{\overline{y}_{k+1}}\sum_{\overline{d}_{k+1}}\sum_{\overline{c}_{k+1}}\sum_{\overline{l}_{k}}y_{k+1}(1-d_{k+1})(1-y_k)\prod_{j=0}^k\frac{I(C_{j+1}=0)}{\Pr[C_{j+1}=0|\overline{L}_j=\overline{l}_j,\overline{C}_j=\overline{Y}_j=\overline{D}_j=0,A=a]}\times\\
&\prod_{j=0}^kf(y_{j+1}|\overline{y}_j,\overline{d}_{j+1},\overline{c}_{j+1},\overline{l}_j,a)f(d_{j+1}|\overline{y}_j,\overline{d}_{j},\overline{c}_{j+1},\overline{l}_j,a)f(c_{j+1}|\overline{y}_j,\overline{d}_{j},\overline{c}_{j},\overline{l}_j,a)\times\\
&f(l_{j}|\overline{y}_j,\overline{d}_{j},\overline{c}_{j},\overline{l}_{j-1},a)=\\
&\sum_{\overline{l}_{k}}\Pr[Y_{k+1}=1|\overline{L}_k=\overline{l}_k,\overline{C}_{k+1}=\overline{Y}_k=\overline{D}_{k+1}=0,A=a]\times\\
&\prod_{j=0}^k\Pr[Y_{j}=0|\overline{L}_{j-1}=\overline{l}_{j-1},\overline{C}_{j}=\overline{Y}_{j-1}=\overline{D}_{j}=0,A=a]\times\\
&\Pr[D_{j+1}=0|\overline{L}_{j-1}=\overline{l}_{j-1},\overline{C}_{j+1}=\overline{Y}_{j}=\overline{D}_{j}=0,A=a]\times\\
&f(l_{j}|\overline{Y}_j=\overline{D}_{j}=\overline{C}_{j}=0,\overline{l}_{j-1},a)
\end{align*}

\begin{customcorollary}{5}\label{corr5}
Suppose \eqref{eqn:C}, \eqref{eqn:positivity}, and \eqref{eqn:consistency} hold.  Then $\Pr[Y^{a,\overline{v}=0}_{k+1}=1|Y^{a,\overline{v}=0}_{k}=0]$ is equivalent to \eqref{eqn:h1v} for any $k=0,\ldots,K$.
\end{customcorollary}
Proof:  By laws of probability we have 
\begin{equation}
\Pr[Y^{a,\overline{v}=0}_{k+1}=1|Y^{a,\overline{v}=0}_{k}=0]=\frac{\Pr[Y^{a,\overline{v}=0}_{k+1}=1,Y^{a,\overline{v}=0}_{k}=0]}{\Pr[Y^{a,\overline{v}=0}_{k}=0]}\label{eqn:writehazard}
\end{equation}
Also
\begin{equation*}
\Pr[Y^{a,\overline{v}=0}_{k+1}=1,Y^{a,\overline{v}=0}_{k}=0]+\Pr[Y^{a,\overline{v}=0}_{k+1}=1,Y^{a,\overline{v}=0}_{k}=1]=\Pr[Y^{a,\overline{v}=0}_{k+1}=1]
\end{equation*}
Because the event $(Y^{a,\overline{v}=0}_{k+1}=1,Y^{a,\overline{v}=0}_{k}=1)$ is equivalent to the event $Y^{a,\overline{v}=0}_{k}=1$ we have
\begin{equation*}
\Pr[Y^{a,\overline{v}=0}_{k+1}=1,Y^{a,\overline{v}=0}_{k}=0]+\Pr[Y^{a,\overline{v}=0}_{k}=1]=\Pr[Y^{a,\overline{v}=0}_{k+1}=1]
\end{equation*}
such that the numerator of \eqref{eqn:writehazard} is equivalent to
\begin{equation}
\Pr[Y^{a,\overline{v}=0}_{k+1}=1,Y^{a,\overline{v}=0}_{k}=0]=\Pr[Y^{a,\overline{v}=0}_{k+1}=1]-\Pr[Y^{a,\overline{v}=0}_{k}=1]\label{eqn:c52}
\end{equation}
which, by Theorem \ref{th1} is equivalent to
\begin{align}
&\Pr[Y^{a,\overline{v}=0}_{k+1}=1]-\Pr[Y^{a,\overline{v}=0}_{k}=1]=\nonumber\\
&\sum_{j=0}^k\sum_{\overline{x}_j}\Pr[Y_{j+1}=1|\overline{X}_j=\overline{x}_j,\overline{V}_{j+1}=\overline{Y}_{j}=0,a]\times\nonumber\\
&\prod_{s=0}^{j}\Pr[Y_{s}=0|\overline{X}_{s-1}=\overline{x}_{s-1},\overline{V}_{s}=\overline{Y}_{s-1}=0,a]f(x_{s}|\overline{x}_{s-1},\overline{V}_{s}=\overline{Y}_{s}=0,a)-\nonumber\\
&\sum_{j=0}^{k-1}\sum_{\overline{x}_j}\Pr[Y_{j+1}=1|\overline{X}_j=\overline{x}_j,\overline{V}_{j+1}=\overline{Y}_{j}=0,a]\times\nonumber\\
&\prod_{s=0}^{j}\Pr[Y_{s}=0|\overline{X}_{s-1}=\overline{x}_{s-1},\overline{V}_{s}=\overline{Y}_{s-1}=0,a]f(x_{s}|\overline{x}_{s-1},\overline{V}_{s}=\overline{Y}_{s}=0,a)=\nonumber\\
&\sum_{\overline{x}_k}\Pr[Y_{k+1}=1|\overline{X}_k=\overline{x}_k,\overline{V}_{k+1}=\overline{Y}_{k}=0,a]\times\nonumber\\
&\prod_{j=0}^{k}\Pr[Y_{j}=0|\overline{X}_{j-1}=\overline{x}_{j-1},\overline{V}_{j}=\overline{Y}_{j-1}=0,a]f(x_{j}|\overline{x}_{j-1},\overline{V}_{j}=\overline{Y}_{j}=0,a)=\nonumber\\
&\mbox{E}[Y_{k+1}(1-Y_{k})W_{k}^V|A=a]\label{eqn:c53}
\end{align}
with the last equality by Lemma \ref{lemma3}.  Also by Theorem  \ref{th1}, the denominator of \eqref{eqn:writehazard} is equivalent to
\begin{align*}
&\sum_{\overline{x}_{k-1}}\Pr[Y_{k}=0|\overline{X}_{k-1}=\overline{x}_{k-1},\overline{V}_{k}=\overline{Y}_{k-1}=0,a]\times\\
&\prod_{j=0}^{k-1}\Pr[Y_{j}=0|\overline{X}_{j-1}=\overline{x}_{j-1},\overline{V}_{j}=\overline{Y}_{j-1}=0,a]f(x_{j}|\overline{x}_{j-1},\overline{V}_{j}=\overline{Y}_{j}=0,a)=\\
&\mbox{E}[(1-Y_{k})W_{k}^V|A=a]
\end{align*}
with the last equality by Lemma \ref{lemma3}.

\begin{customcorollary}{6}\label{corr6}
Given \eqref{eqn:C1}, \eqref{eqn:positivity1}, and \eqref{eqn:consistency1}, the counterfactual hazard under elimination of competing events \eqref{eqn:hazard1} is equivalent to expression \eqref{eqn:h1} for any $k=0,\ldots,K$.
\end{customcorollary}
Proof: The result follows from Corollary \ref{corr5} by choosing $\overline{V}_{K+1}=(\overline{C}_{K+1},\overline{D}_{K+1})$ and $\overline{X}_K=\overline{L}_K$. 

\begin{customcorollary}{7}\label{corr7}
Given \eqref{eqn:C2}, \eqref{eqn:positivity2}, and \eqref{eqn:consistency2}, the counterfactual hazard without elimination of competing events \eqref{eqn:hazard2} is equivalent to expression \eqref{eqn:H2} for any $k=0,\ldots,K$.
\end{customcorollary}
Proof: The result follows from Corollary \ref{corr5} by choosing $\overline{V}_{K+1}=\overline{C}_{K+1}$ and $\overline{X}_K=(\overline{D}_K,\overline{L}_K)$. 

\begin{customcorollary}{8}\label{corr8}
Suppose the following identifying assumptions hold:

\begin{enumerate}
\item Exchangeability:
\begin{equation*}
\overline{D}_{K+1}^{a,\overline{c}=0}\coprod A,
\end{equation*}
\begin{equation}
\underline{D}_{k+1}^{a,\overline{c}=0}\coprod C_{k+1}|%
\overline{L}_k=\overline{l}_k,\overline{D}_k=\overline{C}_k=0,A=a  \label{eqn:Cd}
\end{equation}

\item Positivity:
\begin{align}
&f_{A,\overline{L}_{k},C_k,D_k}(a,\overline{l}_{k},0,0)\neq0\implies\nonumber\\
&\Pr[C_{k+1}=0|\overline{L}_k=\overline{l}_{k},C_k=D_k=0,A=a]>0\mbox{ w.p.1}. \label{eqn:positivityd}
\end{align}

\item Consistency:
\begin{align}
&\mbox{If } A=a \mbox{ and }\overline{C}_{k+1}=0\mbox{,} \nonumber\\
&\mbox{ then } \overline{L}_{k+1}=\overline{L}_{k+1}^{a,\overline{c}=0} \mbox{ and } \overline{D}_{k+1}=\overline{D}_{k+1}^{a,\overline{c}=0} \label{eqn:consistencyd}
\end{align}
\end{enumerate}
allowing $\overline{Y}_k$ to be an implicit component of $\overline{L}_k$.  Then $\Pr [D_{K+1}^{a,\overline{c}=\overline{0}}=1]$, the counterfactual risk of the competing event by $K+1$ under $a$ and elimination of loss to follow-up, is equivalent to 
\begin{align}
\sum_{\overline{l}_{K}}\sum_{k=0}^{K}&\Pr [D_{k+1}=1|\overline{L}_{k}=%
\overline{l}_{k},\overline{C}_{k+1}=\overline{D}%
_{k}=0,A=a]\times  \notag \\
&\prod_{j=0}^{k}\{ \Pr [D_{j}=0|\overline{L}_{j-1}=\overline{l}_{j-1},%
\overline{C}_{j}=\overline{D}_{j-1}=0,A=a]\times  \notag \\
& f(l_{j}|\overline{l}_{j-1},\overline{C}_{j}=\overline{D}%
_{j}=0,a)\} \label{Dgengform}
\end{align}
\end{customcorollary}
Proof: The result follows from Theorem \ref{th1} by replacing $\overline{Y}_{k+1}$, $\overline{V}_{k+1}$, $\overline{X}_k$ and corresponding counterfactuals from Theorem \ref{th1} with $\overline{D}_{k+1}$,$\overline{C}_{k+1}$, $\overline{L}_k$ and their corresponding counterfactuals, respectively.

\underline{Remark}: Note that Corollary \ref{corr8} is agnostic as to whether the event of interest and the competing event are mutually competing events or whether a semi-competing risks setting applies.  In the special case of mutually competing events, separating $\overline{Y}_k$ from $\overline{L}_k$, the g-formula (\ref{Dgengform}) can be more explicitly written as
\begin{align}
\sum_{\overline{l}_{K}}\sum_{k=0}^{K}&\Pr [D_{k+1}=1|\overline{L}_{k}=%
\overline{l}_{k},\overline{C}_{k+1}=\overline{D}%
_{k}=\overline{Y}_k=0,A=a]\times  \notag \\
&\prod_{j=0}^{k}\{ \Pr [D_{j}=0|\overline{L}_{j-1}=\overline{l}_{j-1},%
\overline{C}_{j}=\overline{D}_{j-1}=\overline{Y}_{j-1}=0,A=a]\times  \notag \\
&\Pr [Y_{j}=0|\overline{L}_{j-1}=\overline{l}_{j-1},%
\overline{C}_{j}=\overline{D}_{j}=\overline{Y}_{j-1}=0,A=a]\times  \notag \\
& f(l_{j}|\overline{l}_{j-1},\overline{C}_{j}=\overline{D}%
_{j}=\overline{Y}_j=0,a)\} \label{Dgengformmutually}
\end{align}

\begin{customcorollary}{9}\label{corr9}
Suppose that the assumptions \eqref{eqn:Cd}, \eqref{eqn:positivityd}, \eqref{eqn:consistencyd},  \eqref{eqn:C2}, \eqref{eqn:positivity2} and \eqref{eqn:consistency2} all simultaneously hold.  Then the counterfactual hazard conditioned on competing events \eqref{eqn:hazard3} is equivalent to expression \eqref{eqn:cs1} for any $k=0,\ldots,K$.
\end{customcorollary}
Proof: By probability rules
\begin{equation}
\Pr [Y_{k+1}^{a,\overline{c}=\overline{0}}=1|D_{k+1}^{a,\overline{c}=%
\overline{0}}=Y_{k}^{a,\overline{c}=\overline{0}}=0] =\frac{\Pr [Y_{k+1}^{a,\overline{c}=\overline{0}}=1,D_{k+1}^{a,\overline{c}=%
\overline{0}}=Y_{k}^{a,\overline{c}=\overline{0}}=0] }{\Pr [D_{k+1}^{a,\overline{c}=%
\overline{0}}=Y_{k}^{a,\overline{c}=\overline{0}}=0] }\label{eqn:c92}
\end{equation}
Because the event $Y_{k+1}=1$ implies $D_{k+1}=0$ under any single counterfactual world and by \eqref{eqn:c52} and \eqref{eqn:c53} of Corollary \ref{corr5}, we have that the numerator of \eqref{eqn:c92} is equivalent to 
\begin{align*}
&\Pr [Y_{k+1}^{a,\overline{c}=\overline{0}}=1,D_{k+1}^{a,\overline{c}=%
\overline{0}}=Y_{k}^{a,\overline{c}=\overline{0}}=0]=\\
&\Pr [Y_{k+1}^{a,\overline{c}=\overline{0}}=1,Y_{k}^{a,\overline{c}=\overline{0}}=0]=\\
&\mbox{E}[Y_{k+1}(1-Y_{k})W_{k}^{\overline{c}=0}|A=a]=\\
&\mbox{E}[Y_{k+1}(1-D_{k+1})(1-Y_{k})W_{k}^{\overline{c}=0}|A=a]
\end{align*}
Also, by Theorem \ref{th1} and Corollary \ref{corr2}, replacing the event $Y^{a,\overline{c}=0}_{k+1}=0$ with the joint event $(D^{a,\overline{c}=0}_{k+1}=0,Y^{a,\overline{c}=0}_{k}=0)$, we can write the denominator of \eqref{eqn:c92} as
\begin{align*}
\Pr [D_{k+1}^{a,\overline{c}=%
\overline{0}}=Y_{k}^{a,\overline{c}=\overline{0}}=0]=&\sum_{\overline{l}_k}\prod_{j=0}^{k}\Pr[D_{j+1}=0|\overline{L}_{j}=\overline{l}_{j},\overline{C}_{j+1}=\overline{Y}_{j}=\overline{D}_j=0,a]\\
&\Pr[Y_j=0|\overline{L}_{j-1}=\overline{l}_{j-1}\overline{C}_{j}=\overline{Y}_{j-1}=\overline{D}_j=0,a]f(l_{j}|\overline{l}_{j-1},\overline{C}_{j}=\overline{D}_j=\overline{Y}_{j}=0,a)=\\
&\mbox{E}[(1-D_{k+1})(1-Y_{k})W_{k}^{\overline{c}=0}|A=a]
\end{align*}
with the last equality following arguments similar to Lemma \ref{lemma3}.  

Similar arguments prove the counterfactual hazard of the competing event itself \eqref{eqn:hazard4} is equivalent to expression \eqref{eqn:cs2} given \eqref{eqn:Cd}, \eqref{eqn:positivityd}, \eqref{eqn:consistencyd},  \eqref{eqn:C2}, \eqref{eqn:positivity2} and \eqref{eqn:consistency2}.  

\section{Estimators of the risk of the competing event itself and data application implementation details}\label{dataappendix}
\subsection{Parametric g-formula and IPW estimators of the risk of the competing event}
Parametric g-formula or IPW estimators of the risk of the competing
event $\Pr [D_{K+1}^{a,\overline{c}=\overline{0}}=1]$ as in \eqref{eqn:risk3}
can be obtained using nearly identical algorithms reviewed in Section \ref{totalstat}
when events are mutually competing. Under our
temporal order assumptions within each interval, a parametric g-formula estimator
of the risk of the competing event itself by $K+1$ as identified by (\ref{Dgengformmutually}) and adjusting only for $L_0$
is analogously a function of the estimated observed cause-specific hazards of the event of interest and competing event: 
\begin{equation}
\frac{1}{n}\sum_{i=1}^{n}\sum_{k=0}^{K}q(a,l_{0i},k;\hat{\eta}%
)\prod_{j=0}^{k-1}\left[1-p(a,l_{0i},j;\hat{\theta})\right] \left[%
1-q(a,l_{0i},j;\hat{\eta})\right]  \label{eqn:gcomp2ce}
\end{equation}
Alternatively, an IPW estimator based on weighted estimates of these
cause-specific hazards is 
\begin{equation}
\sum_{k=0}^{K}\hat{h}^{2}_{k}(a;\hat{\alpha})\prod_{j=0}^{k-1}[\{1-\hat{h}%
^{1}_{j}(a;\hat{\alpha})\}\{1-\hat{h}^{2}_{j}(a;\hat{\alpha})\}]
\label{eqn:ipw2estce}
\end{equation}
Finally, an IPW estimator based on estimating observed subdistribution
hazards is as in \eqref{eqn:ipw2firstest} of the main text but treating the original
competing event as the event of interest and the original event of interest
as the competing event. For semi-competing risk settings (e.g. when the
original event of interest is diagnosis of prostate cancer), there are no
competing events for the purposes of estimating the risk of the competing
event . In this case, the original competing event will act as the event of
interest $Y_{k+1}$, time-varying status of the original event of interest may act as a
component of $L_k$ (if it is needed to ensure exchangeability \eqref{eqn:Cd} under the assumptions encoded in the causal diagram) and $\overline{D}_{k+1}\equiv 0$ at all $k$. Here,
methods of Section \ref{directstat} can be used with the competing event status simply set to 0 for all
individuals at all times; in other words, methods for settings where
competing events do not exist can be used.

\subsection{Data application implementation details}
This section gives implementation details for each estimator used in the data analysis described in Section \ref{analysis}.  R code is provided in the supplementary materials.
\subsubsection{Structure of input data sets}
\underline{Data set 1}: The same input data set was used for 1) the parametric g-formula estimators \eqref{eqn:gcomp1} and \eqref{eqn:gcomp2}, 2) the IPW estimators based on cause-specific hazards \eqref{eqn:ipw1est} and \eqref{eqn:ipw2est} for direct and total effects on the event of interest, and 3) the parametric g-formula and IPW estimators \eqref{eqn:gcomp2ce} and \eqref{eqn:ipw2estce} for the total effect on the competing event.  This input data set was constructed as a person-time data set such that each person has $%
K^{*}+1$ lines indexed by $k=0,\ldots,K^{*}$ and measurements of $(L_0,A,C_{k+1},D_{k+1},Y_{k+1})$ on each line $k$. Individuals surviving all causes and not lost to follow-up through the 5-year follow-up have $%
K^{*}+1=60$ records in the data.  An individual experiencing one of these events prior to month $60$ will have $K^{*}+1<60$ records with $K^{*}+1$ the month of this event.   

\underline{Data set 2}: A different input data set was used for the IPW estimator based on subdistribution hazards \eqref{eqn:ipw2firstest} used for the total effect on the event of interest.  Unlike the data set above, this data set keeps individuals in the data set after occurrence of the competing event (i.e. keeps them in the ``risk set'').   This dataset was also constructed as a person-time data set such that each person will have $%
K^{*}+1$ lines indexed by $k=0,\ldots,K^{*}$ and measurements of $(L_0,A,C_{k+1},D_{k+1},Y_{k+1})$ on each line $k$. Individuals surviving all causes and not lost to follow-up through the five-year follow-up \textsl{or} individuals experiencing a competing event within this period have $K^{*}+1=60$ records in the data.  An individual experiencing prostate cancer death or loss to follow-up prior to month $60$ will have $K^{*}+1<60$ records with $K^{*}+1$ the month of this event.  

\underline{Data set 3}: A third input data set was used for the IPW estimator based on subdistribution hazards for the total effect on the competing event.  This was constructed in the same way as data set 2 but treating the original competing event as the event of interest and treating the original event of interest as the competing event.

\subsubsection{Model assumptions}
\underline{For estimators implemented using data set 1:}

We fit a pooled over time logistic regression model for the observed cause-specific hazard of prostate cancer death (the event of interest) in each month $k+1$, $p_{k}(a,l_{0},k;\theta)$, with dependent variable $Y_{k+1}$ and independent variables a third degree polynomial function of month $k+1$, treatment status $A$, interactions between $A$ and the terms for month, and the following function of $L_0$: dichotomized activity level (normal versus reduced activity function), indicators of age group ($\leq59$, $60-75$, $\geq 75$) , dichotomized hemoglobin level ($<12$ versus $\geq12$), and an indicator of previous history of cardiovascular disease.  

We similarly fit a pooled over time logistic regression model for the observed cause-specific hazard of other death (the competing event) in each month $k+1$, $q_{k}(a,l_{0},k;\eta)$ with dependent variable $D_{k+1}$ and independent variables a second degree polynomial function of month, treatment status $A$, and the following function of $L_0$: dichotomized activity level (normal versus reduced activity function), indicators of age group ($\leq59$, $60-75$, $\geq 75$) , dichotomized hemoglobin level ($<12$ versus $\geq12$), and an indicator of previous history of cardiovascular disease.  Note this less flexible model for the observed hazard of other death compared with our model for the observed hazard of prostate cancer death was chosen after several bootstrap samples for the construction of confidence intervals failed to converge under the more flexible model for the other death hazard (using the same independent variables as for the prostate cancer death model).  We did construct point estimates under the more and less flexible model for the other death hazard as a sensitivity analysis which were very similar.

No patient was loss to follow-up in this study prior to $k=50$.  Therefore, we set estimates of the loss to follow-up hazard, $r(a,\overline{l}_{0},k;\hat{\alpha})\equiv r(k)\equiv1$, for all $k<50$ in the weight denominators for all IPW estimators.  For records with index $k\geq 50$, we fit a pooled over time logistic regression model for the loss to follow-up hazard at these times $r(a,\overline{l}_{0},k;\alpha)$ with dependent variable $C_{k+1}$ and independent variables treatment status $A$ and the following function of $L_0$: dichotomized activity level (normal versus reduced activity function), indicators of age group ($\leq59$, $60-75$, $\geq 75$), and an indicator of previous history of cardiovascular disease.  Hemoglobin was originally included in this model but was ultimately excluded in the final analysis as it resulted in failure of model convergence in several bootstrap samples.  We constructed point estimates under a more flexible model that additionally included dichotomized hemoglobin level ($<12$ versus $\geq12$) as a sensitivity analysis and these were similar..

\underline{For estimators implemented using data sets 2 or 3:}
For these estimators, only a model for the loss to follow-up hazard (amongst those previously surviving all causes) is required.   As for estimators based on data set 1, and for individuals still surviving the competing event by $k$, we set this hazard to 1 for all $k<50$ and fit the same pooled over time logistic regression model for the loss to follow-up hazard for $k\geq 50$ $r(a,\overline{l}_{0},k;\alpha)$.  We refit this model in data set 2 (and 3) so programs could be standalone (although the predicted values obtained from data set 1 for the other estimators alternatively could have been saved and used).  To refit these models to data set 2 (or 3), the model fit must be explicitly restricted to records such that failure from the competing event has not yet occurred because these data sets keep failed individuals in the risk set after failure.  For individuals who have failed from the competing event by $k$, we set the loss to follow-up hazard to 1 (because once an individual experiences the competing event without prior censoring, he cannot subsequently be lost to follow-up, as explained in Section \ref{censdef}).

\section{Cross-world alternatives to controlled direct effects }\label{crossworld}
We have focused our attention in this paper on \textsl{single-world}
counterfactual estimands \citep{swigs}. In particular, all of the counterfactual risks and hazards considered in
Section \ref{estimand} could be identified (and, in turn, estimated) in a
study where $A$ is randomized at baseline and censoring is eliminated. Of
course, when censoring events include competing events as in the case of the direct effect \eqref{eqn:RD1}, a controlled direct effect, it is unclear how such a
randomized study could be implemented.  

A \textsl{principal stratum effect}\citep{robinsfail,Frangakisps} is an alternative to a controlled direct effect that quantities a causal effect of the treatment on the event of interest that does not capture the treatment's effect on the competing event.  Specifically, the \textsl{survivor average causal
effect (SACE)} \citep{rubinsace} on the risk of the event of interest by $%
k+1 $ can be written as 
\begin{equation}
\Pr[Y_{k+1}^{a=1,\overline{c}=\overline{0}}=1|D_{k+1}^{a=0,\overline{c}=\overline{0}%
}=0,D_{k+1}^{a=1\overline{c}=\overline{0}}=0]-
\Pr[Y_{k+1}^{a=0,\overline{c}=\overline{0}}=1|D_{k+1}^{a=0,\overline{c}=\overline{0}%
}=0,D_{k+1}^{a=1\overline{c}=\overline{0}}=0]   \label{eqn:sace}
\end{equation}
The estimand \eqref{eqn:sace} is nearly equivalent to the total effect
\eqref{eqn:RD2}. The only distinction is that the effect \eqref{eqn:sace} generally corresponds to a different population than the original study population.  It is restricted to the subset of individuals who would not
experience the competing event by $k+1$ under either level
of treatment $A$. As no individual in this subpopulation will experience a competing event by this time, the effect \eqref{eqn:sace} does not require conceptualizing interventions on the competing event as in the case of 
the controlled direct effect \eqref{eqn:RD1}; in turn, competing events are not censoring events relative to the estimand \eqref{eqn:sace}.

Several authors have given identifying assumptions and corresponding estimators for point estimates or bounds for the SACE \citep{sace,causalitypearl,sace1,sace2,sace3,sace4,sace5,sace6,sace7,sace8,sace9}.  However, a major drawback of this estimand is that, even in cases where such assumptions hold, we do not know which or how large a subset of the population constitutes this principal stratum of survivors even in a study with $A$ randomized and no censoring.  For example, for those with $%
A=1$ we can never observe $D_{k+1}^{a=0,\overline{c}=\overline{0}}$. 

\textsl{Natural direct effects} \citep{rgmediation,pearldirect} constitute another way to define the causal effect of a treatment on an outcome that does not capture the treatment's effect on the mediator.  The natural direct effect of treatment assignment $A$ on the risk of the event of interest by $k+1$ that does not capture the treatment's effect on the competing event (under an intervention that eliminates loss to follow-up) can be defined as 
\begin{equation}
\Pr[Y_{k+1}^{a=1,\overline{c}=\overline{0},\overline{D}^{a=0,\overline{c}=0}_{k+1}}=1] \mbox{ vs.} \Pr[Y_{k+1}^{a=0,\overline{c}=\overline{0},\overline{D}^{a=0,\overline{c}=0}_{k+1}}=1]  \label{eqn:nde}
\end{equation}
The right hand side of \eqref{eqn:nde} is the risk of the event of interest by $k+1$ had we intervened on all individuals in the study population to set $A=0$, eliminate loss to follow-up and set the history of the competing event to the values it would take had we set $A=0$; this is equivalent to the (single-world) risk without elimination of competing events $\Pr[Y_{k+1}^{a,\overline{c}=\overline{0}}=1]$ as in \eqref{eqn:risk2} for $a=0$.   However, the left hand side of \eqref{eqn:nde} is the risk of the event of interest by $k+1$ had we intervened on all individuals in the study population to set $A=1$, eliminate loss to follow-up and set the history of the competing event to the values it would have taken had we set $A=0$.  There is no circumstance in which we can identify the left hand side of \eqref{eqn:nde} without untestable
assumptions because we can never observe $\overline{D}^{a=0,\overline{c}=0}_{K+1}$ once $A$ has been set to 1.  Unlike the controlled direct effect, the natural direct effect has an indirect analogue that sums to the total effect\citep{rgmediation,tylerbook}.  Also, unlike the SACE and like the controlled direct effect, the natural direct effect quantifies an effect in the original (known) study population.  However, like the controlled direct effect, the natural direct effect also requires conceptualizing intervention on competing events but requires even stronger assumptions for identification\citep{rgmediation,tylerbook}.

\underline{Remark:} As we clarified in the main text, when the outcome corresponds to an indicator of failure from an event of interest by $k+1$, a competing event such as death prior to $k+1$ does not result in an undefined value for the outcome but rather a zero value.  Therefore, we can define a total effect in this case.  In alternative settings, where the outcome corresponds to a characteristic of an individual at a particular time that is only defined while he is alive (e.g. blood pressure at time $k+1$), then death prior to $k+1$ renders the outcome undefined such that total treatment effects in the original study population are not defined.  In this setting the term \textsl{truncation by death} is used instead of \textsl{competing events}.   Perhaps due to this inability to define the total effect for outcomes subject to truncation by death, there has been much more explicit debate in the causal inference literature regarding choices of estimand when truncation by death is present \citep{pearlps,joffeps,tylerps}.

\end{document}